\documentclass[12pt]{article}
\pdfoutput=1
\usepackage[a4paper]{geometry}
\usepackage{jheppub, amsmath,amssymb,amsfonts,amsxtra, mathrsfs, makeidx,graphics,graphicx,amsthm,epsfig, bm,longtable,float, color,tikz,mathtools,xfrac,footnote,rotating, lscape, makecell, environ,mathtools, empheq}

\usepackage{subfig}
\usepackage{multirow}
\usepackage{adjustbox}

\usepackage{amsthm}
\usepackage{hyperref}

\usepackage{amsmath}
\usepackage{amsfonts}
\usepackage{commath}
\usepackage[english]{babel}
\usepackage{setspace}

\usepackage{epsfig,amssymb} 
\usepackage{amsmath}

\usepackage{amscd,color}
\usepackage{amsmath,graphicx}
\usepackage{verbatim,booktabs}

\usepackage{latexsym}
\usepackage{amsmath,amsfonts,amssymb,amsthm}
\usepackage{amsmath,amsthm}

\DeclareMathOperator{\Tr}{Tr}

\usepackage{tikz}
\usetikzlibrary{tikzmark}
\usetikzlibrary{math} 
\usetikzlibrary{matrix}
\usetikzlibrary{calc}

\usepackage{array}

\definecolor{mandarancio}{RGB}{255,147,0}
\definecolor{azzurro}{RGB}{0,118,186}
\definecolor{terrabruciatadiconcorezzo}{RGB}{181,23,0}

\def\centerarc[#1](#2)(#3:#4:#5)
{ \draw[#1] ($(#2)+({#5*cos(#3)},{#5*sin(#3)})$) arc (#3:#4:#5) }

\title{
\begin{center} 
3d $\mathcal{N}=2$   SO/USp  adjoint SQCD:\\
s-confinement and exact identities
\end{center}
}

\author[a]{Antonio Amariti}
\author[a,b]{and Simone Rota}
\affiliation[a]{INFN, Sezione di Milano, Via Celoria 16, I-20133 Milano, Italy}

\affiliation[b]{Dipartimento di Fisica, Universit\`a degli Studi di Milano, Via Celoria 16, I-20133 Milano, Italy}
\emailAdd{antonio.amariti@mi.infn.it,  simone.rota@mi.infn.it}

\abstract{
We study 3d $\mathcal{N}=2$ SQCD with symplectic and orthogonal gauge groups and adjoint matter. For $USp(2n)$ with two fundamentals and $SO(N)$ with one vector these models have been recently shown to s-confine.
Here we corroborate the validity of this proposal by relating it to the confinement of $USp(2n)$ with four fundamentals and an antisymmetric tensor, using exact mathematical results coming from the analysis of the partition function on the squashed three-sphere. 
Our analysis allows us to conjecture new s-confining theories for a higher number of fundamentals and vectors, in presence of linear monopole superpotentials. We then prove the new dualities through a chain of  adjoint deconfinements and s-confining dualities.
\\
\\
\\
\emph{To the memory of Luciano Girardello}
}

\begin{document}

\maketitle

\section{Introduction}
\label{sec:intro}

A crucial aspect underlining the study of gauge theories is that gauge invariance corresponds to a redundancy more than to a fundamental symmetry.
This motivates the search of dual models, often described in terms of 
new gauge groups sharing the same IR properties of the original one.
An interesting possibility is that the dual model is described in terms of the confined degrees of freedom of the original one. 
In this case the original model is referred as s-confining 
and it corresponds, in many cases, to a limiting case of a duality between two gauge theories.
Examples of this behavior have been worked out in models preserving four supercharges in 4d and in 3d, namely with $\mathcal{N}=1$ and $\mathcal{N}=2$ supersymmetry respectively.

In the  4d $\mathcal{N}=1$ case with a single gauge group with $W=0$ 
a systematic classification has been proposed by \cite{Csaki:1996sm,Csaki:1996zb}, and elaborating on that results many other examples have been found.
Many examples of this phenomenon in the 3d $\mathcal{N}=2$ case can be 
obtained through the circle compactification of the 4d parent cases, along the lines of \cite{Aharony:2013dha}.

In 3d there is a new ingredient that makes the classification more intricate and offers new examples of gauge theories with confining dynamics, given by the possibility of turning on  monopole superpotentials.
Many examples of  3d s-confining gauge theories been studied  in \cite{Aharony:2013dha,Aharony:2013kma,Csaki:2014cwa,Amariti:2015kha,Amariti:2015xna,Nii:2017npz,Benvenuti:2018bav,Amariti:2018wht,Nii:2018erm,Nii:2018tnd,Nii:2018wwj,Nii:2019dwi,Nii:2019ebv}, where many checks of the new proposed dualities have been performed.
In a recent paper \cite{Benvenuti:2021nwt} models with real gauge groups and adjoint matter 
have been studied and new confining dualities have been proposed.
An interesting aspect of these cases is that the dualities can be proved by sequentially deconfining the adjoint (symmetric or antisymmetric tensors) in terms of other known dualities involving real gauge groups without any tensor.
Such a deconfinement of two-index matter fields follows from the 
one originally worked out in 4d in \cite{Berkooz:1995km} and then refined in \cite{Luty:1996cg} (see also the recent works \cite{Bottini:2022vpy,Bajeot:2022kwt} where  such deconfinement technique has been reconsidered in the 4d case).
In 3d the structure of confining gauge theories is richer 
because of the possibility of turning on monopole superpotentials.

In this paper we elaborate on these results, showing the matching of the three-sphere partition function between the new dual phases proposed by \cite{Benvenuti:2021nwt}.
We find that there is a straightforward proof of the hyperbolic integral identity that corresponds to the matching of the squashed three-sphere partition functions  between the dual phases. The result follows from the identity relating $USp(2 n)$ with the antisymmetric and four fundamentals without monopole superpotential and its description in terms of confined degrees of freedom.
In this case by opportunely fixing the value of the mass parameters and by applying the \emph{duplication formula} for the hyperbolic Gamma functions we observe that the identity can be manipulated into the expected ones for the  new dualities proposed by \cite{Benvenuti:2021nwt}.

This correspondence motivates us to make one step further, and to consider the case of $USp(2 n)$ with the antisymmetric and six fundamentals, in presence of  a monopole superpotential (see  
\cite{Benini:2017dud,Amariti:2017gsm,Benvenuti:2017kud,Benvenuti:2017bpg,Giacomelli:2017vgk,Amariti:2018gdc,Aprile:2018oau,Pasquetti:2019uop,Pasquetti:2019tix,ArabiArdehali:2019zac,Benvenuti:2020gvy,Benvenuti:2021com} for recent examples of 3d $\mathcal{N}=2$ gauge theories and dualities with monopole superpotential turned on).  This model is confining as well and
it admits the same manipulation referred above on the integral identity matching the squashed three-sphere partition functions.
Again we obtain identities relating, in this case,  the partition function of models with $USp(2n)$ or $SO(N)$ gauge groups with four fundamentals or three vectors and an adjoint matter field, and the partition function of models with (interacting) singlets.

We then analyze these models through sequentially deconfining the adjoint fields, obtaining a prove of the dualities. This last approach offers also an alternative derivation of the integral identities (obtained so far through the duplication formula), in terms of adjoint deconfinement. Indeed, as we will explicitly show below, each step discussed in the physical proof of the duality corresponds to the application of a known identity between hyperbolic integrals.

The paper is organized as follows.
In section \ref{sec:rev} we discuss some review material that will be necessary for our analysis. More concretely in sub-section \ref{blm} we review the dualities worked out in \cite{Benvenuti:2021nwt}
while in sub-section \ref{ZS3} we focus on 
 the hyperbolic integrals corresponding to the squashed three-sphere partition function that will play a relevant role in the rest of the paper.
 In section \ref{sec:old} we show how it is possible to reproduce the dualities of \cite{Benvenuti:2021nwt} by an application of the \emph{duplication formula} on the partition function of $Usp(2n)$ with four fundamentals and an antisymmetric.
 Section \ref{sec:new} is the main section of the paper and it contains the new results. Here we start our analysis by reverting the logic discussed so far in the derivation of the dualities. Indeed we first apply the \emph{duplication formula} to the partition function of $Usp(2n)$ with six fundamentals and an antisymmetric. This gives raise to three new integral identities that 
 we interpret as examples of s-confining dualities for 
 $USp(2n)$ or $SO(N)$ gauge theories with four fundamentals or three vectors and an adjoint matter field.
 By flipping some singlets we propone also the structure of the 
 superpotential for the confined phase in each case.
 Then in sub-section \ref{consisency}, as a consistency check, we engineer a real mass flow interpolating from our new dualities to the ones of \cite{Benvenuti:2021nwt}.
 In sub-section \ref{theproof}  we prove 
  the new dualities through deconfining the adjoint matter fields.
  As a bonus we show that this procedure can be followed step by step on the partition function, giving an independent proof of the  integral identities we started with.
  In section \ref{sec:conc} we summarize our analysis and discuss some further lines of research.
In appendix \ref{sec:A} we discuss the physical derivation of the integral identities for the dualities of \cite{Benvenuti:2021nwt} by using the deconfining trick, corroborating the idea of proving exact mathematical identities from physical principles.
In appendix \ref{sec:B} we derive the integral identities for $SO(N)$ gauge theories with $N+1$ vectors and linear monopole superpotential, that have played a prominent role in our analysis.


\section{Review}
\label{sec:rev}

\subsection{3d confining models with real gauge groups and adjoint matter}
\label{blm}
These dualities have been proved in \cite{Benvenuti:2021nwt} and they are the starting point of our analysis.
Here we review the main properties of these dualities and briefly discuss 
their derivation.
Then in appendix \ref{sec:A} we will provide the matching of the three-sphere partition function by reproducing the deconfinement of the adjoint matter fields.

The three classes of s-confining dualities with adjoint matter obtained in \cite{Benvenuti:2021nwt} are summarized in the following.
\begin{itemize}
\item 
In the first case the electric side of the duality involves an 
$USp(2n)$ gauge theory with adjoint $S$ and two fundamentals $p$ and $q$ with superpotential $W = \Tr (p S p)$.
The dual model corresponds of a WZ model with $4n$ chiral multiplets. 
These $4n$ gauge fields corresponds to  gauge invariant 
singlets of the electric theory. There are $2n$ dressed monopole operators, 
$Y_j = Y_{USp} \Tr S^j$, $j=0,\dots,2n-1$, where $Y_{USp}$ is the unit flux monopole of the $USp(2n)$ gauge theory. Then there are $n$ dressed mesons 
$M_\ell = q S^{2 \ell+1} q$ with $\ell=0,\dots,n-1$ and eventually there are $n$ singlets $\sigma_k =\Tr S^{2k}$ with $k=1,\dots,n$.
\item The second case involves an $SO(2n)$ gauge theory with an adjoint $A$ and a vector $q$, without superpotential. The dual theory is a WZ model with $4n$ chiral fields, corresponding to  gauge invariant 
singlets of the electric theory. 
There are $2n-1$ dressed monopole operators, 
$Y_j^+ = Y_{SO}^+\Tr A^j$, $j=0,\dots,2n-2$, where $Y_{SO}^+$ is the unit flux monopole of the $SO(2n)$ gauge theory with positive charge with respect to the charge conjugation symmetry.
Then there are $n$ dressed mesons $M_\ell = q A^{2\ell} q$ with $\ell=0,\dots,n-1$ and $n-1$ singlets $\sigma_k =\Tr A^{2k}$ with $k=1,\dots,n-1$.
The last two chiral fields correspond to the baryon $\mathbb{B} \equiv \text{Pf } A$ and to the  baryon monopole $ Y_{ A^{n-1}}^-$,  obtained from   the unit flux monopole of the $SO(2n)$ gauge theory with negative charge with respect to the charge conjugation symmetry.
\item The third and last case involves an $SO(2n+1)$ gauge theory, again with an adjoint $A$, a vector $q$ and vanishing superpotential. The dual theory is a WZ model with $4n+2$ chiral fields, corresponding to  gauge invariant 
singlets of the electric theory. 
There are $2n$ dressed monopole operators, 
$Y_j^+ = Y_{SO}^+\Tr A^j$, $j=0,\dots,2n-1$, where $Y_{SO}^+$ is the unit flux monopole of the $SO(2n)$ gauge theory with positive charge with respect to the charge conjugation symmetry.
Then there are $n$ dressed mesons $M_\ell = q A^{2\ell} q$ with $\ell=0,\dots,n-1$ and $n$ singlets $\sigma_k =\Tr A^{2k}$ with $k=1,\dots,n$.
The last two chiral fields correspond to the baryon $\mathbb{B} = \epsilon_{2n+1}  (q A^n)$ and to the  baryon monopole $ Y_{q A^{n-1}}^-  $
\end{itemize}

As stressed in \cite{Benvenuti:2021nwt} the superpotential of the dual models correspond to polynomials of the singlets and with complexity that rapidly grows when the ranks of the gauge groups increase.
Nevertheless by flipping the singlets $\sigma_k$, and the baryon  and the baryon monopole in the orthogonal cases, these superpotentials are given by cubic combinations of the remaining singlets.

Let us briefly sketch the strategy for proving these dualities. The first step consists of  deconfining the adjoint field. In the symplectic case the adjoint is in the symmetric representation and it can be deconfined in terms of an orthogonal gauge group. On the other hand in the orthogonal case the adjoint is in the antisymmetric representation and it can be deconfined in terms of a symplectic gauge group. 
In each case this step requires to find a confining duality that reduces to the original model.  
After deconfining the adjoint one is then left with a two gauge node quiver gauge theory and one can then proceed by dualizing the original gauge node, by using a known duality.
In the cases at hand this duality corresponds to a limiting case of an
Aharony duality or a modification of it, with monopole superpotentials.
This gives raise to another model with a real gauge group and adjoint matter and generically  a more sophisticated superpotential. By repeating the procedure of rank-two tensor deconfinement and duality one is left with the original gauge group but with rank of one unit less and it allows to iterate the procedure 
and arrive to the desired WZ model at the end of such a cascading process.

By inspection it has been shown in \cite{Benvenuti:2021nwt}  that the adjoint of the $USp(2n)$ case 
can be deconfined by an $SO(2n+1)$ gauge group and a superpotential flipping the monopole. After dualizing the $USp(2n)$ gauge theory one ends up with an $SO(2n+1)$ gauge theory with an adjoint and a dynamically generated superpotentials flipping both the monopole and the baryon monopole. In this case the adjoint can be deconfined by an $USp(2n-2)$ 
gauge group and a more intricate flavor structure. Indeed  the $SO(2n+1)$/$USp(2n-2)$ gauge group have one extra vector/fundamental charged chiral fields and there is a superpotential interactions between these two fields and the $SO(2n+1) \times USp(2n-2)$ bifundamental.
Furthermore there is a linear monopole superpotential for the $USp(2n)$ gauge node.
By dualizing the $SO(2n+1)$ gauge node with $2n$ vectors one ends up with an $USp(2n-2)$ gauge theory, with two fundamentals and a non trivial superpotential. By opportunely flipping some of the singlets of the original model one can  recast that the original $USp(2n)$, iterate the procedure and eventually prove the duality.
Similar analysis have been used to prove the orthogonal dualities as well.
In such cases after deconfining the antisymmetric in terms of $USp(2n-2)$ and dualizing the original orthogonal gauge group one is left with $USp(2n-2)$ and two fundamentals. Then the duality proven above for this case can be used to prove the duality for the orthogonal cases as well.

\subsection{Confining theories and the three-sphere partition function}
\label{ZS3}

Here we review some known aspect of the 3d partition function for 3d 
$\mathcal{N}=2$  gauge theories on the squashed three-sphere preserving 
$U(1) \times U(1)$ isometry.

The real  squashing parameter $b$ can be associated to two imaginary parameters $\omega_1 = i b$ and $\omega_2 = i/b$ and their combination is usually referred as $2 \omega \equiv \omega_1+\omega_2$.
The matter and the vector multiplets contribute to the partition function through hyperbolic Gamma function, defined as
\begin{equation}
  \label{Gamma}
  \Gamma_h(x;\omega_1,\omega_2) \equiv
  \Gamma_h(x)\equiv 
  e^{
    \frac{\pi i }{2 \omega_1 \omega_2}
    \left((x-\omega)^2 - \frac{\omega_1^2+\omega_2^2}{12}\right)}
  \prod_{j=0}^{\infty} 
  \frac
  {1-e^{\frac{2 \pi i}{\omega_1}(\omega_2-x)} e^{\frac{2 \pi i j \omega_2 }{\omega_1}}}
  {1-e^{-\frac{2 \pi i}{\omega_2} x} e^{-\frac{2 \pi i j \omega_1 }{\omega_2}}}.
\end{equation}
The argument $x$ represents a parameters associated to the real scalar
in the (background) vector multiplet and it gives the informations about the representations and the global charges of the various fields.
We refer the reader to \cite{Benvenuti:2021nwt} for further details.

Here we are interested in two confining gauge  with $USp(2n)$ gauge group and antisymmetric and six or four fundamentals. In the first case the theory has a monopole superpotential and it corresponds to the reduction of a 4d
$\mathcal{N}=1$ confining gauge theory. In the second case the theory with four fundamenrtals can be obtained by a real mass flow, it is still confining but in this case the superpotential is vanishing.
Details on these models have been discussed in \cite{Amariti:2018wht,Benvenuti:2018bav}.

In general the partition function of an $USp(2n)$ gauge theory with $2n_f$ fundamentals and an antysimmetric tensor is
\begin{eqnarray}
Z_{\tau,\vec \mu}^{USp(2n)}
=
\frac{ \Gamma_h(\tau)^n }
{(-\omega_1 \omega_2)^\frac{n}{2} 2^{n} n!}
\int
\prod_{a=1}^{n} dy_a  
\frac{ \prod_{r=1}^{2n_f}\Gamma_h (\pm y_a+ \mu_r) }{\Gamma_h(\pm 2 y_a)}
\prod_{1\leq a<b \leq n}\frac{ \Gamma_h(\pm y_a \pm y_b+\tau )}{ \Gamma_h(\pm y_a \pm y_b)} \nonumber \\
\end{eqnarray}
Where the parameters $\tau$ and $\mu_r$ are associated to the antisymmetric tensor and to the $2n_f$ fundamentals respectively.
The two confining dualities discussed above for $2n_f=6$ and $2n_f=4$ correspond to the following  identities
\begin{equation}
\label{bc}
Z_{\tau,\mu_1\dots,\mu_6}^{USp(2n)} = 
 \prod_{j=0}^{n-1} \Gamma_h((j+1)\tau)
  \prod_{1\leq r<s\leq 6}
\Gamma_h(j \tau+\mu_r+\mu_s)
\end{equation}
with the balancing condition
\begin{equation}
\label{bbbccc}
2(n-1) \tau +\sum_{a=1}^{6} \mu_a = 2 \omega
\end{equation}
signaling the presence of a linear monopole superpotential, and
\begin{equation}
\label{initial}
Z_{\tau, \mu_1,\dots,\mu_4}^{USp(2n)} = 
 \prod_{j=0}^{n-1} \frac{\Gamma_h((j+1)\tau)}{\Gamma_h((2n-2-j)\tau+\sum_{r=1}^{4}\mu_r)}
 \prod_{1\leq r<s\leq 4}
\Gamma_h(j \tau+\mu_r+\mu_s)
\end{equation}
with unconstrained parameters, corresponding to the absence of 
any monopole superpotential.

These identities are the starting point of our analysis, and they contain
all the mathematical information on the models with real gauge groups and adjoint matter.

In order to transform symplectic gauge groups into unitary one we will use 
a well known trick, already used in the literature \cite{Dolan:2008qi,Spiridonov:2010qv,Spiridonov:2011hf,Benini:2011mf}. It consists of  using the duplication formula \cite{doi:10.1063/1.531809,10.5555/1075051.1716652,+2003+839+876}
\begin{equation}
\label{duplication}
\Gamma_h(2x) = \Gamma_h(x) \Gamma_h\left(x+\frac{\omega_1}{2}\right)  \Gamma_h\left(x+\frac{\omega_2}{2}\right)  \Gamma_h(x+\omega) 
\end{equation}
to modify the partition function of the vector multiplet of $USp(2n)$ 
into the  partition function of the vector multiplet of $SO(2n)$ or $SO(2n+1)$.

This transformation requires to consider an $USp(2n)$ gauge theory with fundamental matter fields and assign to some of the mass parameters some specific value as $\mu =\pm \frac {\omega_i}{2}$ or $\mu = \omega$ or $\mu=0$. Then by applying the duplication formula (and the reflection  equation 
$\Gamma(x) \Gamma(2 \omega-x)=1$ when necessary) one can convert the 
contribution of $USp(2n)$ with fundamentals in the one of $SO(2n)$ or $SO(2n+1)$ with (few) vectors.
Furthermore, by using the same mechanism, one can convert also the contribution of the $USp(2n)$ antisymmetric field into the one of an adjoint (for both the symplectic and the orthogonal cases).

To simplify the reading of the various steps of the derivation we  conclude this section by  summarizing  the integral identities for $USp(2n)$ and $SO(N)$  s-confining SQCD, that we have used in the analysis below.
These identities are indeed necessary for translating into the language of the squashed three-sphere partition function
 the chain of adjoint deconfiments and dualities introduced above. In the table we indicate the gauge group, the matter content, the superpotential and the reference to the integral identity equating the partition function of each gauge theory with the one of  its confined description . 
\begin{center}
\begin{tabular}{||c|c|c|c||}
\hline
Gauge group & Matter & Superpotential &Identity \\
\hline
$USp(2n)$ & $2n+4 \, \square$ & $W=Y_{USp}$ & (\ref{USp}) \\
$USp(2n)$ & $2n+2 \, \square$ & $W=0$& (\ref{eq:aharony_USp}) \\
\hline
$SO(2n)$ & $2n+1 \, \square$ & $W=Y_{SO}+$ & (\ref{SOeven}) \\
$SO(2n)$ & $2n-1 \, \square$ & $W=0$& (\ref{eq:aharony_SO_even_2}) \\
\hline 
$SO(2n+1)$ & $2n+2 \, \square$ &$W=Y_{SO}+$ & (\ref{SOdd}) \\
$SO(2n+1)$ & $2n \, \square$ & $W=0$& (\ref{eq:aharony_SO_odd_2}) \\
\hline
\end{tabular}
\end{center}

\section{Proving known results}
\label{sec:old}

In this section we show how to obtain the integral identities for the three dualities reviewed in subsection (\ref{blm}) by applying the duplication formula (\ref{duplication}) on the identity (\ref{initial}).
Here and in the following section we will use three choice of masses, that are
\begin{center}
\begin{tabular}{cccc}
I. & $\vec \mu_{n_f} $ & $=$ & $ \left(\frac{\tau}{2}+\frac{\omega_1}{2},\frac{\tau}{2}+\frac{\omega_2}{2},\frac{\tau}{2},\vec \mu_{n_f-3}\right)$
\\
II. &  $\vec \mu_{n_f}$&$ = $&$\left(\frac{\omega_1}{2},\frac{\omega_2}{2},0,\vec \mu_{n_f-3}\right)$
\\
III. &  $\vec \mu_{n_f} $&$=$&$ \left(\frac{\omega_1}{2},\frac{\omega_2}{2},\tau, \vec \mu_{n_f-3} \right)$
\end{tabular}
\end{center}
Here we did not specify the length $n_f$ of the vector $\vec \mu$. 
In the following we will have $n_f=4$ for the cases of \cite{Benvenuti:2021nwt} and $n_f=6$
for the new dualities discussed here.  

\subsection*{Case I: $USp(2n)$}

If we choose the  masses  $\mu_r$ as $\vec \mu = \left(\frac{\tau}{2}+\frac{\omega_1}{2},\frac{\tau}{2}+\frac{\omega_2}{2},\frac{\tau}{2},m\right)$ and apply the duplication formula, the LHS of  (\ref{initial}) becomes
\begin{eqnarray}
\label{firstsp}
&&
\frac{\Gamma_h(\tau)^n}{(-\omega_1 \omega_2)^{\frac{n}{2}}2^n n!}
\int_{C^n}\prod_{1\leq j<k \leq n} \frac{\Gamma_h(\tau\pm x_j \pm x_k)}{\Gamma_h(\pm x_j \pm x_k)}
 \nonumber \\
 \times 
 &&
 \prod_{j=1}^{n}  \frac{\Gamma_h(\tau\pm 2x_j) 
 \Gamma_h(m\pm x_j)\Gamma_h(\omega-\frac{\tau}{2}\pm x_j)}{\Gamma_h(\pm 2x_j)} d x_j  
\end{eqnarray} 
This corresponds to the partition function of $USp(2n)$ with an adjoint $S$, a fundamental $p$ and a fundamental $q$ with superpotential 
$W = Tr(p S p)$, where the constraint imposed by the superpotential corresponds to the presence of the parameter $\omega-\frac{\tau}{2}$ in the argument of the last hyperbolic gamma function in the numerator of (\ref{firstsp}).

On the other hand the RHS  of (\ref{initial})  requires more care.
Let us separate first the contributions of the three terms.
By substituting the parameters $\mu_r$ and using the reflection
equation we have
\begin{eqnarray} 
&&
\prod_{j=0}^{n-1} 
\Gamma_h(\omega-\big(2n-j-\frac{1}{2} \big)\tau) -m)
 \nonumber \\
\times &&
\Gamma_h((j+1)\tau, (j+1) \tau+\frac{\omega_1}{2},
(j+1) \tau+\frac{\omega_2}{2},
(j+1) \tau+\omega) \nonumber \\
\times &&
\Gamma_h(\big(j+\frac{1}{2} \big) \tau+\frac{\omega_1}{2} + m,
\big(j+\frac{1}{2} \big) \tau+\frac{\omega_2}{2} + m,
\big(j+\frac{1}{2} \big) \tau + m \big)
\end{eqnarray}
where we used the shorthand notation $\Gamma_h(a,b) = \Gamma_h(a) \Gamma_h(b)$.
By using the duplication formula it becomes
\begin{eqnarray} 
\prod_{j=0}^{n-1} 
\frac{\Gamma_h\left(\omega-\big(2n-j-\frac{1}{2} \big)\tau) -m,
2(j+1)\tau,(2j+1)\tau + 2m\right)}{
\Gamma_h(\big(j+\frac{1}{2} \big) \tau+m+\omega)}
\end{eqnarray}
This last formula  can be reorganized as
\begin{eqnarray} 
\label{lastuspblm}
&&
\prod_{j=0}^{2n-1} 
\Gamma_h\left(\omega-\big(2n-j-\frac{1}{2} \big)\tau -m\right)
\cdot
\prod_{\ell=0}^{n-1} 
\Gamma_h((2\ell+1)\tau + 2m)
\cdot
\prod_{k=1}^{n} 
\Gamma_h(2 k \tau)
\nonumber \\
\end{eqnarray}
The three terms in the argument of these hyperbolic Gamma function correspond to the ones expected from the duality. 
Indeed if we associate a mass parameter $\tau$ to the adjoint and two mass parameters $m_1=m$ and $m_2 = \omega-\frac{\tau}{2}$ then the unit flux bare monopole $Y_{Usp}$ has mass parameter 
$m_{Y_{Usp} }= 2\omega - 2n\tau-m_1-m_2$. The dressed monopole $Y_j=Y_{Usp}  S^j $ has  mass parameter $m_{Y_j} =2\omega - (2n-j)\tau-m_1-m_2$. By using the constraint imposed by the superpotential on $m_2$ we then arrive at 
$m_{Y_j} =\omega - (2n-j-\frac{1}{2})\tau-m$, corresponding to the argument of the first hyperbolic Gamma function in (\ref{lastuspblm}).
On the other hand the arguments of the second and of the third Gamma functions in (\ref{lastuspblm}) are straightforward and they correspond to the dressed mesons $M_\ell = q S^{2\ell+1} q$ and the to the singlets 
$\sigma_k =\Tr S^{2k}$.
\subsection*{Case II: $SO(2n)$}
In this case we choose the parameters  $\mu_r$ as  $\vec \mu = \left(\frac{\omega_1}{2},\frac{\omega_2}{2},0,m\right)$ and apply the
  duplication formula.  On the LHS of (\ref{initial}) we obtain
\begin{eqnarray}
&&
\frac{\Gamma_h(\tau)^n}{(-\omega_1 \omega_2)^{\frac{n}{2}}2^n n!}
\int_{C^n}\prod_{1\leq j<k \leq n} \frac{\Gamma_h(\tau\pm x_j \pm x_k)}{\Gamma_h(\pm x_j \pm x_k)}
 \prod_{j=1}^{n}  \Gamma_h(m\pm x_j) d x_j  
\end{eqnarray} 
This corresponds to the partition function of $SO(2n)$ with an adjoint $A$ and a vector $q$ with vanishing superpotential.
Actually to correctly reproduce the expected partition function we need an extra factor of $2$, in order to have $2^{n-1}$ in the denominator, that correctly reproduces the Weyl factor. This extra $2$ will be generated when  looking at the RHS as are going to explain.

The  RHS  of (\ref{initial}) can be studied as in the $USp(2n)$ case above.
In this case we obtain 
\begin{eqnarray}
\label{risSO}
&&
\frac{1}{2} \Gamma_h(n \tau) \cdot
\Gamma_h(\omega-(n-1)\tau -m) \cdot
\prod_{k=1}^{n-1} \Gamma_h(2 k \tau)
\nonumber \\
\times &&
\prod_{\ell=0}^{n-1} \Gamma_h(2 \ell \tau+2m) \cdot
\prod_{j=0}^{2n-2} \Gamma_h(\omega-(2n-2-j) \tau-m)
\end{eqnarray}
where we used the duplication formula, the reflection
equation and the relations 
$\Gamma_h
\left(
\frac{\omega_1}{2}
\right)=\Gamma_h
\left(
\frac{\omega_2}{2}
\right)=
\frac{1}{\sqrt 2}$.
As anticipated above, the $\frac{1}{2}$ term can be moved on the LHS reproducing the Weyl factor of $SO(2n)$. The other contributions correspond to the $4n$ singlets of \cite{Benvenuti:2021nwt}
.
Let us discuss them in detail.
Again we associate a mass parameter $\tau$ to the adjoint and a mass parameters $m$ to the vector.
The unit flux bare monopole $Y_{SO}^+$ has mass parameter 
$m_{Y_{SO}^+} = \omega - 2(n-1)\tau-m$. The dressed monopoles $Y^+_j = Y_{SO}^+\Tr A^j$ have mass parameter $m_{Y^+_j} =\omega - (2n-2-j)\tau-m$, corresponding to the last term in the second line of (\ref{risSO}).
The mass parameter associated to the  baryon monopole is obtained by adding $(n-1) \tau $ to $m_{Y_{SO}^+}$. This gives  $m_{Y_{A^{n-1}}^-} = \omega - (n-1)\tau-m$ and it corresponds  to the second term in the first line of  (\ref{risSO}). The first term of  (\ref{risSO}), with
mass parameter $n \tau$ corresponds to the baryon $\mathbb{B} \equiv \text{Pf } A$. 
The dressed mesons $M_\ell = q A^{2\ell} q$ and the singlets $\sigma_k =\Tr A^{2k}$ are associated to the combinations $m_{M_\ell } =2 \ell \tau+2m$ and $m_{\sigma_k}=2 k \tau$   respectively.
\subsection*{Case III: $SO(2n+1)$}
 In this case we choose the parameters  $\mu_r$ as  $\vec \mu = \left(\frac{\omega_1}{2},\frac{\omega_2}{2},\tau,m\right)$ and apply the
  duplication formula.  On the LHS of (\ref{initial}) we obtain
\begin{eqnarray}
&&
\frac{\Gamma_h(\tau)^n}{(-\omega_1 \omega_2)^{\frac{n}{2}}2^n n!}
\int_{C^n}\prod_{1\leq j<k \leq n} \frac{\Gamma_h(\tau\pm x_j \pm x_k)}{\Gamma_h(\pm x_j \pm x_k)}
 \prod_{j=1}^{n} \frac{ \Gamma_h(\tau \pm x_j) \Gamma_h(m\pm x_j)}{\Gamma_h(\pm x_j)} d x_j  
 \nonumber \\
\end{eqnarray} 
This corresponds to the partition function of $SO(2n+1)$ with an adjoint $A$ and a vector $q$ with vanishing superpotential. Actually we are still missing a contribution $\Gamma_h(m)$ coming from the zero modes of the vector.
As in the $SO(2n)$ case discussed above, the extra term comes from the RHS, that in this case becomes
\begin{eqnarray}
\label{risSOd}
&&
\frac{\Gamma_h(\omega-n \tau) \Gamma_h(n \tau+m)}{\Gamma_{h}(m)}
\prod_{k=1}^{n} \Gamma_h(2 k \tau) \prod_{\ell=0}^{n-1} \Gamma_h(2 \ell \tau+2m)
\nonumber \\
\times &&
\prod_{j=0}^{2n-1} \Gamma_h(\omega-(2n-1-j) \tau-m)
\end{eqnarray}
As anticipated above the denominator can be moved on the LHS and 
it is necessary to reproduce the zero mode of the chiral fields in the vectorial representation of the $SO(2n+1)$ gauge group.
The other $4n+2$ Gamma functions correspond to the singlets discussed 
in \cite{Benvenuti:2021nwt}.
Let us discuss them in detail.
Again we associate a mass parameter $\tau$ to the adjoint and a mass parameters $m$ to the vector.
The unit flux bare monopole $Y_{SO}^+$ has mass parameter 
$m_{Y_{SO}^+} = \omega - (2n-1)\tau-m$. The dressed monopoles $Y^+_j = Y_{SO}^+\Tr A^j$ have mass parameter $m_{Y^+_j} =\omega - (2n-1-j)\tau-m$, corresponding to the  term in the second line of (\ref{risSO}).
The baryon monopole $ Y_{q A^{n-1}}^- $ is obtained by adding $(n-1) \tau +m $ to the contribution of $m_{Y_{SO}^+} $. This gives
$m_{Y_{q A^{n-1}}^-} = \omega - (n-1)\tau$, and this gives raise to the first term in the first line of  (\ref{risSOd}). The second term in the first line of  (\ref{risSOd}), with mass parameter $n \tau+m$ corresponds to the baryon $\epsilon_{2n+1}  (q A^n)$. 
The dressed mesons $M_\ell = q A^{2\ell} q$ and the singlets $\sigma_k =\Tr A^{2k}$ are associated to the combinations $m_{M_\ell } =2 \ell \tau+2m$ and $m_{\sigma_k}=2 k \tau$   respectively.

\section{New results}
\label{sec:new}

In this section we propose three  new dualities, that generalize the ones reviewed above, in presence of two more fundamentals (or vectors) and of a monopole superpotential.

Here we propose such dualities by reversing the procedure adopted so far.
We start from the integral identity  (\ref{bc}) , that has a clear physical interpretation, because it gives the mathematical version of  the confinement of $USp(2n)$ with an antisymmetric, six fundamentals and the monopole superpotential.

Then we use the duplication formula and we obtain three new relations as discussed above in terms of $USp(2n)$ ($SO(N)$) with an adjoint $S$ ($A$), four (three) fundamentals (vectors) and  $W=p S p$ ($W=0$).
In each case the masses are constrained because the choice of parameters
necessary to apply the duplication formula leaves us with a constraint, corresponding to the leftover of (\ref{bbbccc}).

 By applying the three choices of mass parameters discussed in Section \ref{sec:old}  we arrive at the following three identities
 
\subsection*{Case I: $USp(2n)$}
The first choice corresponds to choosing  $\vec \mu = \left(\frac{\tau}{2}+\frac{\omega_1}{2},\frac{\tau}{2}+\frac{\omega_2}{2},\frac{\tau}{2},\mu_1,\mu_2,\mu_3\right)$.
Substituting in (\ref{bc}) 
it gives raise to the following identity
\begin{eqnarray}
\label{SPP}
&&
\frac{\Gamma_h(\tau)^n}{(-\omega_1 \omega_2)^{\frac{n}{2}}2^n n!}
\int_{C^n}\prod_{1\leq j<k \leq n} \frac{\Gamma_h(\tau\pm x_j \pm x_k)}{\Gamma_h(\pm x_j \pm x_k)}
 \prod_{j=1}^{n}  \frac{ \Gamma_h(\tau \pm 2x_j)\prod_{r=1}^{4} \Gamma_h(\mu_r\pm x_j)}{\Gamma_h(\pm 2x_j)} d x_j 
 \nonumber \\
 =&&
 \prod_{k=1}^{n} \Gamma_h(2 k \tau)
 \cdot
 \prod_{j=0}^{n-1}  \prod_{r=1}^{3} \Gamma_h((2j+1) \tau + 2 \mu_r)
 \cdot
 \prod_{j=0}^{2n-1}   \prod_{1\leq r<s\leq 3}
\Gamma_h(j \tau+\mu_r+\mu_s)
  \\
 =&&
 \prod_{k=1}^{n} \Gamma_h(2k \tau)
 \cdot
 \prod_{j=0}^{n-1}  \bigg(
  \prod_{1\leq r \leq s\leq 3}
\Gamma_h((2j+1) \tau+\mu_r+\mu_s)
\cdot \prod_{1\leq r<s\leq 3}
\Gamma_h(2j \tau+\mu_r+\mu_s)
\bigg)
\nonumber
\end{eqnarray} 
with the conditions
\begin{equation}
\label{original}
2n \tau +\sum_{a=1}^{4} \mu_a = 2 \omega \quad \& \quad
2\mu_4 +  \tau = 2 \omega
\end{equation}
Schematically this corresponds to:
\begin{equation}
\begin{gathered}
S p(2n) \text{ w/ adjoint S} \\
\text { and } 4 \text { fundamentals } q_{1,2,3}, p\\
W =  Y_{USp}+ \Tr (p S p)\\
\end{gathered}
\quad\Longleftrightarrow\quad
\begin{gathered}
\label{Wusp31}
	\text{Wess-Zumino w/ 10n chirals }\\
	\sigma_k = $ Tr$S^{2k}, \quad k=1,\dots , n\\
	\mathcal{A}_{rs}^{(2\ell)} \equiv q_r S^{2\ell} q_s, \quad r < s \\
	\mathcal{S}_{rs}^{(2\ell+1)}\equiv q_r S^{2\ell+1} q_s ,\quad r\leq s \end{gathered}
\end{equation}
where $ \ell=0,\dots n-1$ and $r,s=1,2,3$.
The dual (confined) model corresponds to a set of singlets, $\sigma_k = $ Tr$S^{2k}$, with $k=1,\dots,n$, and  dressed mesons.
These are in the antisymmetric and in the symmetric representation of the flavor symmetry group that rotates $q_{1,2,3}$ and they can be defined as  $\mathcal{A}_{rs}^{(2 \ell)} 
\equiv q_r S^{(2\ell)} q_s$  and $\mathcal{S}_{rs}^{(2\ell+1)}\equiv q_r S^{2\ell+1} q_s$ respectively.
By flipping the singlets $\sigma_{k}$ we modify the electric theory, adding the superpotential terms $\Delta_{W_{ele}} = \sum_{k=1}^{n} \rho_k\Tr S^{2k}$.
In the dual theory we are left with the cubic superpotential 
\begin{eqnarray}
\label{Wusp}
W &=& \sum_{\ell_1+\ell_2+\ell_3 = 2n-2} 
\epsilon^{r_1 r_2 r_3} 
\epsilon^{s_1 s_2 s_3} 
\mathcal{S}_{s_1,r_1}^{(2\ell_1+1)}  \mathcal{S}_{s_2,r_2}^{(2\ell_2+1)}  \mathcal{S}_{s_3,r_3}^{(2\ell_3+1)}  
\nonumber \\
&+&
 \sum_{\ell_1+\ell_2+\ell_3 = 2n-1} 
\epsilon^{r_1 r_2 r_3} 
\epsilon^{s_1 s_2 s_3} 
\mathcal{A}_{s_1,r_1}^{(2\ell_1)}  \mathcal{A}_{s_2,r_2}^{(2\ell_2)}  \mathcal{S}_{s_3,r_3}^{(2\ell_3+1)}  
\end{eqnarray}
On the identity (\ref{SPP}) the effect of such a flip corresponds to moving the terms $\Gamma_h(2 k \tau)$ on the LHS and taking them to the numerator by  using the reflection
equation, giving raise to the contribution $\Gamma_{h} (2 \omega-2k \tau)$, corresponding to the singlets $\rho_k$.

\subsection*{Case II: $SO(2n)$}

The second choice corresponds to choosing  $\vec \mu = \left(\frac{\omega_1}{2},\frac{\omega_2}{2},0,\mu_1,\mu_2,\mu_3\right)$.
Substituting in (\ref{bc})  gives raise to the following identity
\begin{eqnarray}
\label{SOev}
&&
\frac{\Gamma_h(\tau)^n}{(-\omega_1 \omega_2)^{\frac{n}{2}}2^{n-1} n!}
\int_{C^n}\prod_{1\leq j<k \leq n} \frac{\Gamma_h(\tau\pm x_j \pm x_k)}{\Gamma_h(\pm x_j \pm x_k)}
 \prod_{j=1}^{n}  \prod_{r=1}^{3} \Gamma_h(\mu_r\pm x_j) d x_j 
=
 \Gamma_h(n \tau) 
    \nonumber \\
  && \prod_{k=1}^{n-1} \Gamma_h(2k  \tau)
 \prod_{j=0}^{n-1}  \prod_{r=1}^{3} \Gamma_h(2j \tau + 2 \mu_r)
 \prod_{1 \leq r<s \leq 3}  \Gamma_h((n-1)\tau+\mu_r+\mu_s)
 \prod_{j=0}^{2n-2}  \Gamma_h(j \tau+\mu_r+\mu_s)
 \nonumber \\
\end{eqnarray} 
with the condition
\begin{equation}
\label{BCSO}
2(n-1) \tau +\sum_{r=1}^{3} \mu_r =  \omega
\end{equation}
This corresponds to the duality:
\begin{equation}
\begin{gathered}
	SO(2n) \text{ w/ adjoint A}\\
	\text{and 3 vectors } q_{1,2,3}\\
	 W=Y_{SO}^+\\
\end{gathered}
\quad\Longleftrightarrow\quad
\begin{gathered}
	\text{Wess-Zumino w/ $10n+3$ chirals}\\
	\sigma_k = \Tr A^{2k} \\
	\mathcal{A}_{rs}^{(2\ell+1)} \equiv q_r A^{2\ell+1} q_s \quad  r < s \\
	\mathcal{S}_{rs}^{(2\ell)}\equiv q_r A^{2\ell} q_s  \quad  r \leq s \\
	\mathbb{B} = \text{Pf } A\\
	\mathbb{B}_r = \epsilon_{rst} \, \epsilon_{2n} (A^{n-1} q_s q_t )
\end{gathered}
\end{equation}
with $k=1,\dots , n-1$, $\ell=0,\dots,n-1$ and $r,s=1,2,3$.
The dual description consists of a set of chiral fields identified with mesons and baryons of the electric theory.
The baryon  $\mathbb{B} = \text{Pf } A$ is reproduced on the partition function by $\Gamma_h( n \tau)$
while the baryons $\mathbb{B}_r = \epsilon_{rst} \, \epsilon_{2n} (A^{n-1} q_s q_t )$  are reproduced on the partition function by $\Gamma_h((n-1)\tau+\mu_r+\mu_s)$.
There is also a tower of singlets $\sigma_k$ associated to the singlets $\Tr A^{2k}$ contributing to the partition function as $\prod_{k=1}^{n-1} \Gamma_h(2 k \tau)$.

The mesons are in the antisymmetric and in the symmetric representation of the flavor symmetry group that rotates  the three vectors and they can be defined as  $\mathcal{A}_{rs}^{(2\ell+1)} \equiv q_r A^{2\ell+1} q_s$  and $\mathcal{S}_{rs}^{(2\ell)}\equiv q_r A^{2\ell} q_s$ respectively.
By flipping the  singlets $\sigma_{k}$ and the baryons  we are left, in the dual theory, with the cubic superpotential 
\begin{eqnarray}
\label{Weven}
W &=& \sum_{\ell_1+\ell_2+\ell_3 = 2n-2} 
\epsilon^{r_1 r_2 r_3} 
\epsilon^{s_1 s_2 s_3} 
\mathcal{S}_{s_1,r_1}^{(2\ell_1)}  \mathcal{S}_{s_2,r_2}^{(2\ell_2)}  \mathcal{S}_{s_3,r_3}^{(2\ell_3)}  
\nonumber \\
&+&
 \sum_{\ell_1+\ell_2+\ell_3 = 2n-3} 
\epsilon^{r_1 r_2 r_3} 
\epsilon^{s_1 s_2 s_3} 
\mathcal{A}_{s_1,r_1}^{(2\ell_1+1)}  \mathcal{A}_{s_2,r_2}^{(2\ell_2+1)}  \mathcal{S}_{s_3,r_3}^{(2\ell_3)}  
\end{eqnarray}
Again we can reproduce the effect of the flip on the partition function by
moving the relative Gamma function on the LHS of (\ref{SOev}) and using the reflection
equation.

\subsection*{$SO(2n+1)$}

The third choice corresponds to choosing  $\vec \mu = \left(\frac{\omega_1}{2},\frac{\omega_2}{2},\tau,\mu_1,\mu_2,\mu_3\right)$.
Substituting in (\ref{bc})  gives raise to the following identity
\begin{eqnarray}
\label{SOddd} 
&& 
\frac{\Gamma_h(\tau)^n \prod_{r=1}^{3} \Gamma_h(\mu_r)}{(-\omega_1 \omega_2)^{\frac{n}{2}}2^n n!}
\int_{C^n}\prod_{1\leq j<k \leq n} \frac{\Gamma_h(\tau\pm x_j \pm x_k)}{\Gamma_h(\pm x_j \pm x_k)}
 \nonumber \\
 \times&&
 \prod_{j=1}^{n}  \frac{ \Gamma_h(\tau \pm x_j)\prod_{r=1}^{3} \Gamma_h(\mu_r\pm x_j)}{\Gamma_h(\pm x_j)} d x_j 
 =
 \Gamma_h(\omega-n \tau) \prod_{r=1}^{3} \Gamma_h(n \tau+\mu_r)
 \nonumber \\
 \times &&
 \prod_{k=1}^{n} \Gamma_h(2 k \tau)
 \prod_{j=0}^{n-1}  \prod_{r=1}^{3} \Gamma_h(2j \tau + 2 \mu_r)
 \prod_{j=0}^{2n-1}   \prod_{1\leq r<s\leq 3}
\Gamma_h(j \tau+\mu_r+\mu_s)
 \nonumber \\
\end{eqnarray} 
with the condition
\begin{equation}
\label{constrodd}
(2n-1) \tau +\sum_{a=1}^{3} \mu_a =  \omega
\end{equation}
This corresponds to:
\begin{equation}
\begin{gathered}
	SO(2n+1) \text{ w/ adjoint A}\\
	\text{and 3 vectors } q_{1,2,3}\\
	 W=Y_{SO}^+\\
\end{gathered}
\quad\Longleftrightarrow\quad
\begin{gathered}
	\text{Wess-Zumino w/ $10n+4$ chirals}\\
\sigma_k = \Tr A^{2k} \\
	\mathcal{A}_{rs}^{(2\ell+1)} \equiv q_r A^{2\ell+1} q_s \quad  r < s \\
	\mathcal{S}_{rs}^{(2\ell)}\equiv q_r A^{2\ell} q_s  \quad  r \leq s \\
	\mathbb{B} \equiv \epsilon_{2n+1} A^{n-1} q_1 q_2 q_3\\
	\mathbb{B}_i \equiv \epsilon_{2n+1} A^{n} q_i
\end{gathered}
\end{equation}
with $k=1,\dots , n-1$, $\ell=0,\dots,n$ and $r,s=1,2,3$.
The dual description consists of a set of chiral fields identified with symmetric and antisymmetric mesons as above, the
baryons $\mathbb{B} \equiv \epsilon_{2n+1} A^{n-1} q_1 q_2 q_3 $ and 
$\mathbb{B}_i \equiv \epsilon_{2n+1} A^{n} q_i$ and the singlets 
$\sigma_k = \Tr A^{2k}$.
On the partition function such fields correspond to $\Gamma_h(\omega-n \tau)$, $ \Gamma_h(n \tau+\mu_r)$ and $\prod_{k=1}^{n} 
\Gamma_h(2 k \tau)$ respectively.
Again by  flipping the singlets and leaving only the mesons on the 
dual side we are left with the superpotential  (\ref{Weven}).
We can reproduce the effect of such flip on the partition function by
moving the relative Gamma function on the LHS of (\ref{SOddd}) and using the reflection
equation.
%
%
%
%
\subsection{A consistency check: flowing to the cases of \cite{Benvenuti:2021nwt}}
\label{consisency}
%
%
%
%
Here we show that by giving large masses to two of the fundamentals (or two of the vectors in the theories with orthogonal group) the dualities \eqref{SPP}, \eqref{SOev} and \eqref{SOddd} reduce respectively to the dualities (5.1), (5.2) and (5.3) of \cite{Benvenuti:2021nwt}.

\subsection*{Case I: $USp(2n)$}
We consider the real mass flow triggered by giving large real masses (of opposite signs) to two of the quarks, say $q_1$ and $q_2$. On the electric side we are left with a $USp(2n)$ theory with two quarks $q=q_3$ and $p$, one adjoint and $W=pSp$. The linear monopole superpotential is lifted in the mass flow. 
On the magnetic side the dressed mesons $\mathcal{A}_{3r}^{(2\ell)}$, $\mathcal{S}_{3r}^{(2\ell+1)}$ and $\mathcal{S}_{rr}^{(2\ell+1)}$ with $r=1,2$ become massive and are integrated out in the IR. The dressed mesons $\mathcal{A}_{12}^{(2\ell)}$ and $\mathcal{S}_{12}^{(2\ell+1)}$ are massless and are identified with the dressed monopoles  $Y_j$ of the electric theory. 
More precisely we identify $\mathcal{A}_{12}^{(2 \ell)}$ with $Y_{2\ell}$ and $\mathcal{S}_{12}^{(2\ell+1)}$ with $Y_{2\ell+1}$ for $\ell=0,\dots,n-1$.
The leftover dressed mesons $\mathcal{S}_{33}^{2\ell+1}$ correspond to $M_\ell$, for $\ell=0,\dots,n-1$.
The superpotential (\ref{Wusp}) reduces to the one of \cite{Benvenuti:2021nwt} when the singlets
$\sigma_k$ are flipped.
Indeed the only superpotential terms surviving the real mass flow are
\begin{eqnarray}
\label{WuspBLM}
W &\propto& \sum_{\ell_1+\ell_2+ \ell = 2n-2} 
S_{1,2}^{(2\ell_1+1)}  S_{1,2}^{(2\ell_2+1)}  S_{3,3}^{(2 \ell+1)}  +
 \sum_{\ell_1+\ell_2+ \ell = 2n-1} 
A_{1,2}^{(2\ell_1)}  A_{1,2}^{(2\ell_2)}  S_{3,3}^{(2\ell+1)}  
\nonumber \\
&=&
 \sum_{j_1,j_2,\ell} Y_{{j_1}} Y_{{j_2}} M_\ell \delta_{j_1+j_2+2\ell-4n+2}
\end{eqnarray}

We can follow this real mass flow on the partition function in the following way.
We parametrize the mass parameters as:
\begin{equation}
	\mu_1 = \nu+s, \qquad \mu_2 = \nu-s, \qquad \mu_3 = m
\end{equation}
and we take the limit $s \to \infty$. The constraint from the monopole superpotential reads:
\begin{equation}	\label{eq:constr_USp_flow}
	2\nu = \omega - 2n\tau +\frac{\tau}{2} - m
\end{equation}
On the RHS of \eqref{SPP} the Gamma functions with finite argument in the $s\to\infty$ limit are:
\begin{equation}
\begin{split}
	&\prod_{\ell=0}^{n-1} \Gamma_h \left( (2\ell+1)\tau +2m \right)
	\prod_{\ell=0}^{2n-1} \Gamma_h \left( \ell\tau+2 \nu\right) 
	\\
	&=\prod_{\ell=0}^{n-1} \Gamma_h \left( (2\ell+1)\tau +2 m \right)
	\prod_{j=0}^{2n-1} \Gamma_h \left( \omega+j\tau - 2n\tau +\frac{\tau}{2} - m \right) 
\end{split}
\end{equation}
which correspond to the singlets $M_\ell$ and $Y_j$. On the LHS it corresponds to the partition function of  $USp(2n)$  with 2 fundamentals $p, q$, one adjoint $S$, $n$ singlets $\rho_k$ and superpotential $W=\sum_{k=1}^{n} \rho_k \Tr S^{2k} + p S p$ as expected. The Gamma functions with divergent argument can be written as an exponential using the formula:
\begin{equation}
	\lim_{z\to\pm\infty}  \Gamma_h (z) = \zeta^{-\text{sgn} (z)} 
			\text{exp} \left( \frac{i \pi} { 2 \omega_1 \omega_2} \text{sgn} (z) (z-\omega)^2 \right)
\end{equation}
where $\zeta = \text{exp} \left( 2\pi i \frac{\omega_1^2 + \omega_2^2}{48 \omega_1 \omega_2} \right)$.
The resulting phase on the LHS is then (we omit the prefactor $ \frac{i \pi} { 2 \omega_1 \omega_2}$):
\begin{equation}
	\sum_{j=1}^{n} (s + \nu \pm x_j - \omega)^2 - (-s+\nu \pm x_j - \omega)^2 
	=
	8 s n (\nu - \omega) 
\end{equation}
while on the RHS it is:
\begin{equation}
\begin{split}
	&\sum_{\ell=1}^{n} \left( ((2\ell-1) \tau +2 \nu +2 s-\omega)^2 - ((2\ell-1) \tau + 2\nu -2 s-\omega)^2 \right)
	\\
	&+
	\sum_{\ell=0}^{2n-1} \left( ( \ell \tau + m + \nu + s-\omega)^2 - (\ell \tau + m + \nu - s-\omega)^2 \right)
	\\&=
	4 n s (6\nu +2m -4\omega +(4n-1)\tau)
\end{split}
\end{equation}
Under the constraint \eqref{eq:constr_USp_flow} the divergent phases cancel between the RHS and the LHS. We are then left with an equation which corresponds to the identity between the partition functions of the theories of the duality (5.1) of  \cite{Benvenuti:2021nwt}.

\subsection*{Case II: $SO(2n)$}

We can flow from the duality \eqref{SOev} to (5.2) of  \cite{Benvenuti:2021nwt} by giving a large mass of opposite sign to two vectors.
Indeed the only mesons that survive the projection are the ones labeled by $\mathcal{A}_{12}^{(2 \ell+1)}$, $\mathcal{S}_{12}^{(2\ell)}$ and 
$\mathcal{S}_{33}^{(2\ell)}$.
The first two  are associated to the dressed monopoles $Y_j^+$ as
$\mathcal{A}_{12}^{(2 \ell+1)} =Y_{2\ell+1}^+$ and $\mathcal{S}_{12}^{(2\ell)}=Y_{2\ell}^+$ for $\ell=0,\dots,n-1$.
The leftover dressed mesons $\mathcal{S}_{33}^{2\ell}$ correspond to $M_\ell$.
After the real mass flow the superpotential (\ref{Weven}) reduces to the one of \cite{Benvenuti:2021nwt} when the singlets $\sigma_k$, $Y_{A^{n-1}}^-$ and $Y_{j}^+$  are flipped:
\begin{eqnarray}
\label{WevenBLM}
W &\propto& \sum_{\ell_1+\ell_2+ \ell = 2n-2} 
\mathcal{S}_{1,2}^{(2\ell_1)}  \mathcal{S}_{1,2}^{(2 \ell_2)}  \mathcal{S}_{3,3}^{(2 \ell)}  +
 \sum_{\ell_1+\ell_2+ \ell = 2n-3} 
  \mathcal{A}_{1,2}^{(2\ell_1+1)}  \mathcal{A}_{1,2}^{(2\ell_2+1)}  \mathcal{S}_{3,3}^{(2\ell)}  
\nonumber \\
&=&
 \sum_{j_1,j_2,\ell} Y_{{j_1}}^+ Y_{{j_2}}^+ M_\ell \delta_{j_1+j_2+2\ell-4n+4}
\end{eqnarray}
In order to follow the real mass flow on the partition function we parametrize the masses as:
\begin{equation} 	\label{eq:large_mass}
	\mu_1 = \nu+s, \qquad \mu_2 = \nu-s, \qquad \mu_3 = m
\end{equation}
The constraint reads:
\begin{equation}
	2(n-1)\tau + 2\nu +m = \omega
\end{equation}
Taking the limit $s\to\infty$ the LHS becomes the partition function for $SO(2n)$ with one vector and one adjoint multiplied by a divergent phase. The singlets on the RHS of \eqref{SOev} that remain massless are:
\begin{equation}
\begin{split}
	&\Gamma_h \left(n\tau\right)
	\prod_{k=1}^{n-1} \Gamma_h \left( 2 k \tau \right)
	\prod_{\ell=1}^{n-1} \Gamma_h \left( 2\ell \tau + 2 m \right)
	\Gamma_h \left( \omega - (n-1)\tau -m \right) 
	\\ &\times
	\prod_{j=0}^{2n-1} \Gamma_h \left( \omega +j\tau -m - 2(n-1)\tau \right)
\end{split}
\end{equation}
which correspond respectively to the singlets 
$\mathbb{B}$, $\sigma_k$, $M_\ell$, $Y_{A^{n-1}}^-$ and $Y_{j}^+$ discussed above. Along the lines of the computation done in the previous case one can show that the divergent phases cancel between the LHS and the RHS. The limit $s\to\infty$ then gives the identity between the partition functions of the dual theories (5.2) of  \cite{Benvenuti:2021nwt}.

\subsection*{Case III: $SO(2n+1)$}
When we give large masses to two of the vectors this duality reduces to the duality (5.3) of  \cite{Benvenuti:2021nwt}.
Analogously to the $SO(2n)$ case the  the superpotential
reduces to the one of \cite{Benvenuti:2021nwt} when the singlets
$\sigma_k$, $Y^+_j$ and $Y_{q A^{n-1}}^-$ are flipped.\\

We parametrize the real masses as in \eqref{eq:large_mass}. The constraint reads:
\begin{equation}
	(2n-1)\tau + 2 \nu + m = \omega
\end{equation}
The LHS becomes the partition function for a $SO(2n+1)$ gauge theory with one vector $q$ and one adjoint  $A$  multiplied by a divergent phase. The singlets on the RHS of \eqref{SOddd} that remain massless are:
\begin{equation}
\begin{split}
	&\Gamma_h\left(\omega-n\tau\right)
	\Gamma_h\left(n\tau+m\right)
	\prod_{k=1}^{n} \Gamma_h\left(2 k \tau\right) 
	\prod_{\ell=0}^{n-1} \Gamma_h\left(2 \ell \tau+2m\right)
	\\&\times
	\prod_{j=0}^{2n-1}\Gamma_h\left(\omega + j\tau -m - (2n-1)\tau \right)
\end{split}
\end{equation}
which correspond respectively to the singlets
$Y_{q A^{n-1}}^-$, $\mathbb{B}$, $\sigma_k$ ,
$M_\ell$ and $Y^+_j$
discussed above.
The divergent phases cancel between the LHS and the RHS. The resulting identity corresponds to the duality (5.3) of  \cite{Benvenuti:2021nwt}.

%
%
%
\subsection{Proving the new dualities through adjoint deconfinement}
\begin{figure}
\centering
\includegraphics[width=12cm]{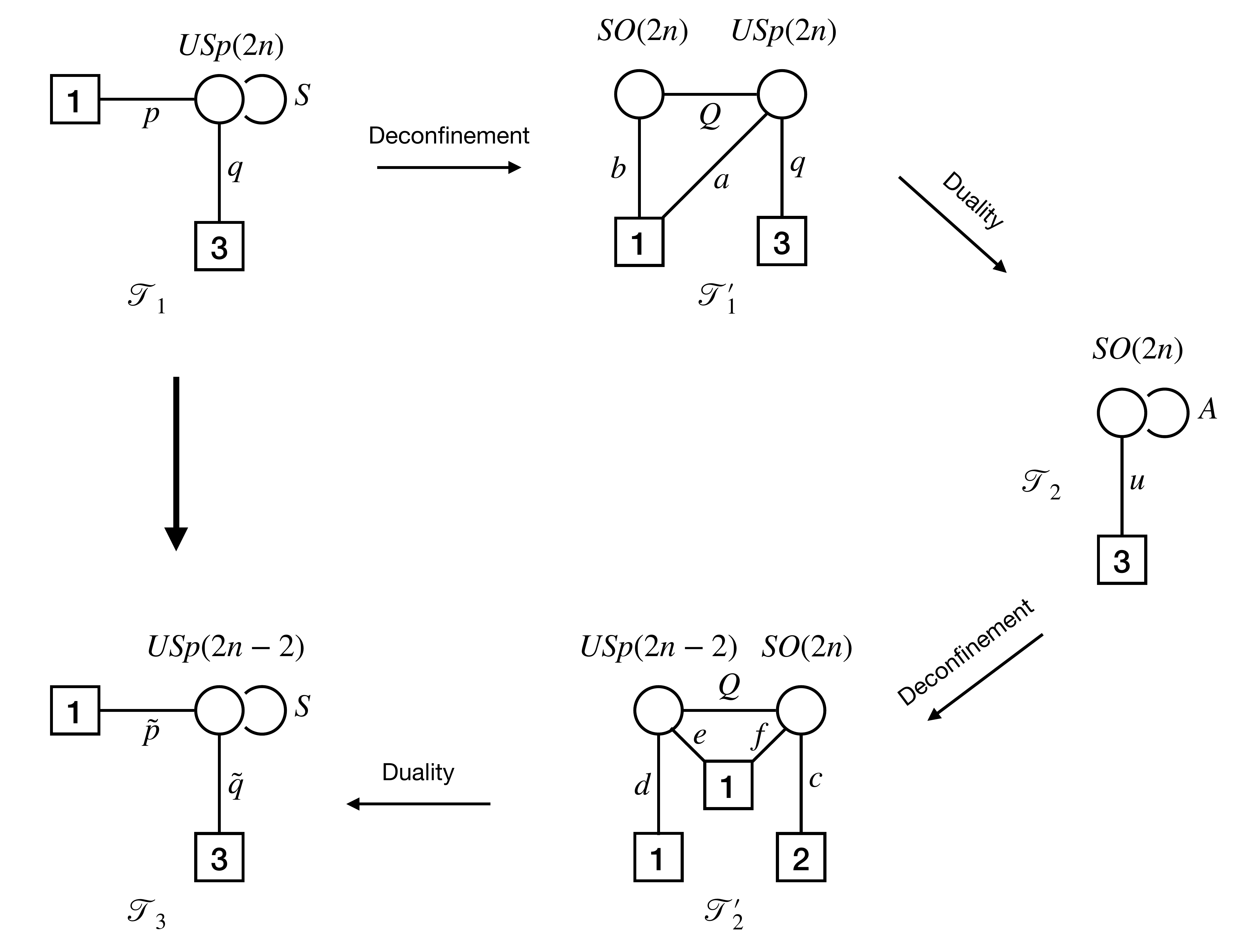}
\caption{ Schematic representation of one step of the deconfinement procedure used to prove the confinement of the $USp(2n)$ model with monopole superpotential. The superpotential and three-sphere partition function of each model in Figure are:}
\vspace{10pt}
\begin{tabular}{||c||c|c|c|c|c||}
\hline
	&$ \mathcal{T}_1$	&	$ \mathcal{T}_1	'$	&$ \mathcal{T}_2$	&$ \mathcal{T}_2'$	&$ \mathcal{T}_3$ \\ \hline
W & \eqref{Wusp31} & \eqref{eq:W_T1prime} &\eqref{SOSOSO} & \eqref{eq:W_T2prime} & \eqref{final}\\
\hline
$\mathcal{Z}_{S^3}$ & \eqref{SPP} &\eqref{eq:Z_T1prime} &\eqref{eq:SO(2n)_step2} & \eqref{eq:Z_T2prime} &\eqref{eq:Z_final}\\
\hline
\end{tabular}
\label{fig:quivers}
\end{figure}

%
%
%
\label{theproof}
The dualities read above from the matching of the three-sphere partition functions can be proved along the lines of \cite{Benvenuti:2021nwt} by deconfining the adjoints
as reviewed in sub-section \ref{blm}.
Even if the logic is very similar the presence of more fundamentals/vectors and the constraints imposed by the monopole superpotentials modify the analysis and it is worth to study explicitly the mechanism. Furthermore 
when translated to the three-sphere partition function this process offers an alternative derivation of the mathematical identities (\ref{SPP}), (\ref{SOev}) and (\ref{SOddd}) from a physical perspective. In Figure \ref{fig:quivers} we show schematically the confinement/deconfinement procedure we used to prove the confinement of the $USp(2n)$ model with monopole superpotential.

%
%
\subsubsection*{Case I: $USp(2n)$}
%
%

\begin{figure}
\begin{center}
\includegraphics[width=10cm]{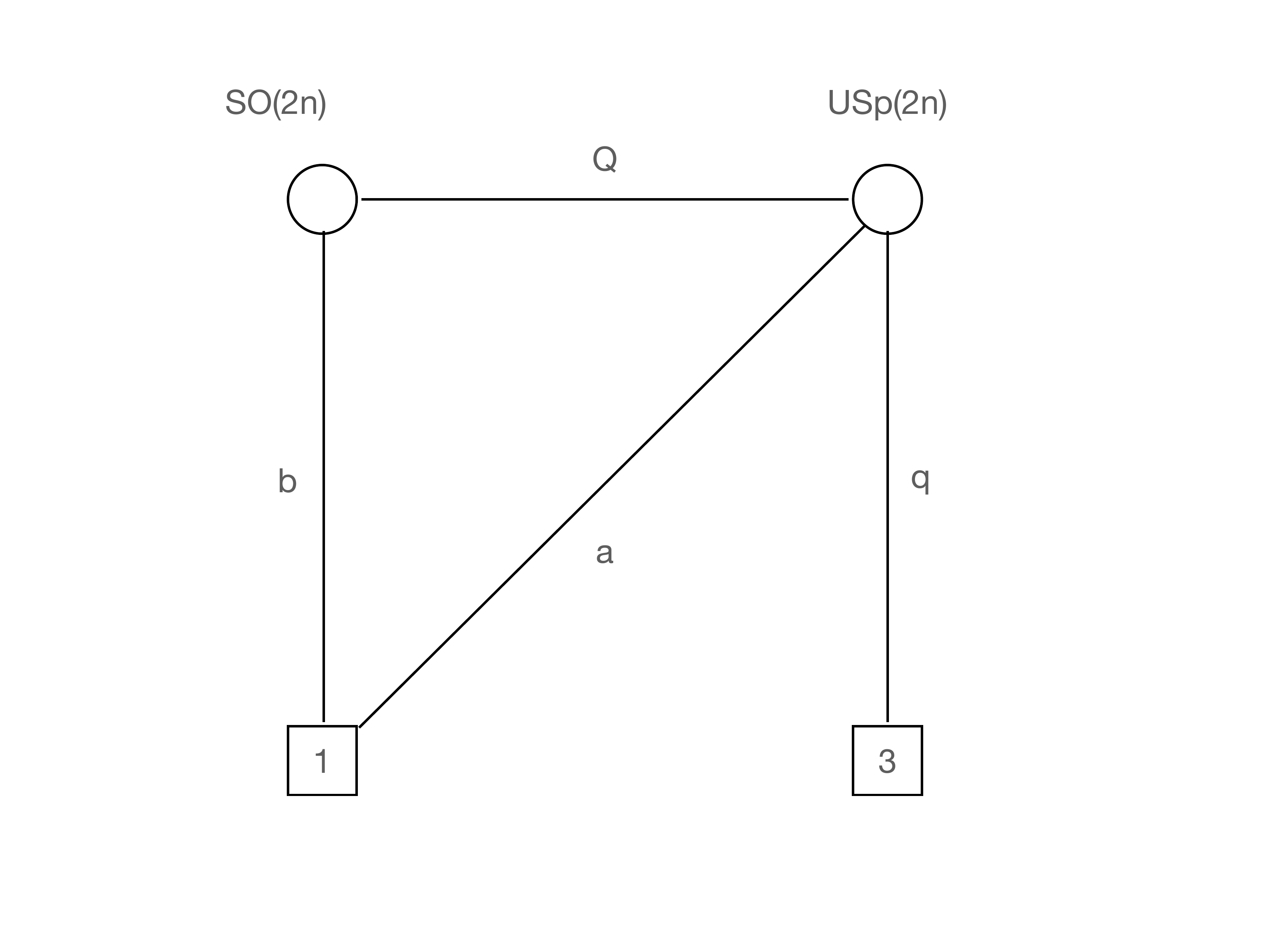}
\caption{Quiver description of the  deconfinement of the adjoint $S$ of
the $USp(2n)$ model with superpotential  (\ref{Wusp31}). In this and in the following quivers we decided to omit to represent the various singlets.}
\label{fig1}
\end{center}
\end{figure}

The $USp(2n)$ model with an  adjoint $S$,  four fundamentals 
$\{q_{1,2,3},p \}$ and superpotential (\ref{Wusp31}) is dual
to the $USp(2n) \times SO(2n)$ quiver given in Figure \ref{fig1}.
As discussed above the analysis is made  easier by  flipping the singlets $\Tr S^{2k}$ with $k=1,\dots,n$.  On the physical side this corresponds to adding singlets $\rho_k$ to the original $USp$ theory with superpotential:
\begin{equation}
\label{deltaWUSp}
	\delta W = \sum_{k=1}^{n} \rho_k\Tr \left(S^{2k}\right)
\end{equation}	
while mathematically it corresponds to moving the tower 
$\Gamma_h(2 k \tau)$ on the LHS of (\ref{SPP}) and by using the reflection
equation we are left with $\Gamma_h(2\omega-2 k \tau)$.
The superpotential associated to the quiver in Figure \ref{fig1} is then\footnote{In $\epsilon_{2n} Q^{2n} $ the $USp$ indices of $Q$ are contracted using $J=  \left(\begin{array}{cc}0 & \mathbb{I}_n \\-\mathbb{I}_n & 0\end{array}\right)$ and the $SO$ indices are contracted with $\epsilon_{2n}$, explicitly $\epsilon_{2n} Q^{2n} = \epsilon_{i_1 j_1 \dots i_n j_n} Q_{i_1}^{a_1} J^{a_1 b_1} Q_{j_1}^{b_n} \dots Q_{i_n}^{a_n} J^{a_n b_n} Q_{j_n}^{b_n}$. Similarly $\Tr \big( S^n \big)$ is a shorthand notation for  $\Tr \big( (S\cdot J)^n \big)$. In the rest of the paper we omit the matrix $J$, which is always understood whenever we contract the indices of a symplectic group.}:
\begin{equation}
\label{eq:W_T1prime}
W = Y_{USp} + Y_{SO}^+ +\Tr (ab Q) + s_1 \epsilon_{2n} ( Q^{2n}) 
+\sum_{k=1}^{n-1} \rho_k Tr\left((QQ)^{2k}\right)
\end{equation}
Indeed by confining the $SO(2n)$ gauge node of this quiver we 
arrive at the original model. This can be proved thanks to a confining duality reviewed in the appendix \ref{sec:B}.  
By confining the $SO(2n)$ node  the superpotential becomes
\begin{eqnarray}
W &=& Y_{USp} +\Tr (a M_{bQ}) 
+ \sum_{k=1}^{n-1} \rho_k\Tr\left(S^{2k}\right)
+ s_1q_b
+p S  p
\nonumber \\
&+&  M_{bb} q_b^2 
+ M_{bQ} q_b p
+\det \left(
\begin{array}{cc}
S & M_{bQ} \\
M_{bQ} & M_{bb}
\end{array}
\right)
\end{eqnarray}
where the duality map for the meson of $SO(N)$ is
\begin{equation}
 \left(
\begin{array}{cc}
Q Q& b Q \\
b Q & bb
\end{array}
\right)
\equiv
 \left(
\begin{array}{cc}
S & M_{bQ} \\
M_{bQ} & M_{bb}
\end{array}
\right)
\end{equation}
while the baryons are $q_b \equiv   \epsilon_{2n} (Q^{2n}) $ 
and $p \equiv  \epsilon_{2n}  (Q^{2n-1} b)$.
 The field $S$ is in the adjoint  of $USp(2n)$ while the field $p$ is  the fourth fundamental of $USp(2n)$ (the other three fundamentals $q_{1,2,3}$ are spectator  when confining $SO(2n)$).
By evaluating the F-terms of the massive fields  we end up with the original $USp(2n)$ gauge theory, with an adjoint, four fundamentals and
superpotential 
\begin{equation}
W = Y_{USp} +p S p + \sum_{k=1}^{n-1} \rho_k\Tr \big(S^{2k}\big)+ M_{bb} \text{det}(S)
\end{equation}
 We can expand the determinant of $S$ in terms of traces\footnotemark:
\begin{equation}
	\text{det} \left(S\right) \propto \Tr\left(S^{2n}\right) + \text{multi-traces}
\end{equation}
By dropping the multi-trace terms and by comparing with the superpotential  $W =$ (\ref{Wusp31}) $+$ (\ref{deltaWUSp})  of the $USp(2n)$ model we started with, we identify $M_{bb} = \rho_n$.

\footnotetext{
	For a $2n\times 2n$ symmetric matrix $S$:
	\begin{equation}
		\text{det} (S)= \text{det} (S\cdot J) = \frac{1}{(2n)!} B_{2n} (s_1, \dots, s_{2n}),
		\qquad
		s_k = (-1)^{k-1} (k-1)!\Tr \left( (S\cdot J)^k \right)
	\end{equation}
	where $B_n$ are Bell polynomials $B_l (s_1, \dots, s_l) = s_l + \dots$.
}

On the partition function the mass parameters for the fields appearing in this 
$USp(2n) \times SO(2n)$ quiver are related to the ones of the original $USp(2n)$ model (i.e. $\mu_r$ and $\tau$ in formula (\ref{SPP}))  by the following set of relations 
\begin{eqnarray}
\sum_{r=1}^{3} \mu_{r} + 2 n \mu_Q +\mu_a = 2 \omega, \quad 2n \mu_Q + \mu_b  = \omega, \quad
\mu_Q + \mu_b + \mu_a  = 2, \quad
\mu_{s_1} + 2 n  \mu_{Q} = 2 
\nonumber \\
\end{eqnarray}
where $\mu_r$ are the three mass parameter for the fields $q_{1,2,3}$.
Furthermore we can map these parameters to the ones in  the confined $SO(2n)$ model by imposing $\mu_Q = \frac{\tau}{2}$.
In this way we arrive at the following identifications
\begin{eqnarray}
\mu_{s_1} = 2 \omega-n \tau,\quad
 \mu_b = \omega-n \tau,\quad
 \mu_a = 2 \omega -\sum_{r=1}^{3} \mu_r -n \tau
\end{eqnarray}
with the constraint
\begin{equation}
2n\tau-\frac{\tau}{2} + \sum_{r=1}^{3} \mu_r = \omega
\end{equation}
The duality between the original $USp(2n)$ model and the quiver with the deconfined  adjoint  can be checked on the partition function by using the identity (\ref{SOeven}). 
This can be shown explicitly by  considering the partition function of the quiver, i.e. 
\begin{eqnarray}
Z_{USp(2n) \times SO(2n)}
&=&
\frac{ \prod_{k=1}^{n-1} \Gamma_h(2 \omega-2k \tau) 
 \Gamma_h(\mu_{s_1})}
{(-\omega_1 \omega_2)^n 2^{2n-1} (n!)^2}
\int 
\prod_{i=1}^{n} dx_i 
\frac{\Gamma_h (\pm x_i + \mu_b) }{ \Gamma_h(\pm x_i)} 
\nonumber \\
&\times&
\prod_{\alpha =1}^{n} dy_\alpha  
 \frac{\Gamma_h (\pm y_\alpha +\mu_a) \prod_{r=1}^3 \Gamma_h (\pm y_\alpha +\mu_r) }{\Gamma_h(\pm 2 y_\alpha)}
\nonumber \\
&\times&
\frac{
\prod_{i=1}^{n}\prod_{\alpha=1}^{n} \Gamma_h \left(\pm x_i \pm y_\alpha +\mu_Q \right)}
{\prod_{i<j} 
\Gamma_h(\pm x_i \pm x_j) \prod_{\alpha<\beta} \Gamma_h(\pm y_\alpha \pm y_\beta)}
\label{eq:Z_T1prime}
\end{eqnarray}
and then by  using the relation  (\ref{SOeven}). This is possible because the  mass parameters of the $2n+1$ vectors in the $SO(2n)$ model are related, due to the linear monopole superpotential, by 
\begin{equation}
\sum_{\alpha=1}^{n} (\pm y_\alpha + \mu_Q) + \mu_b = 2 n \mu_Q + \mu_b = \omega
\end{equation}
By applying  (\ref{SOeven}) and by using the reflection
equation we end up with the first line of  (\ref{SPP}), finding the expected result.

Next we can dualize the $USp(2n)$ node with the linear monopole superpotential turned on. We are left with an $SO(2n)$ SQCD with an adjoint $A$ and superpotential 
\begin{equation}
\label{SOSOSO}
W = Y_{SO}^+ + \sum_{k=1}^{n-1} \rho_k\Tr A^{2k}  + \epsilon_{rst}  
(M_{r s}  v_{t} Pf A+v_{r} \epsilon (A^{n-1} u_{s} u_{t}))
+s_1 Pf A
\end{equation}
In this case the fields are mapped to the ones in the $USp(2n) \times SO(2n)$ quiver as $u_r = Q q_r$,  $A = Q Q$,   $M_{rs} = q_r q_s$ and $v_r= a q_s$.
The fields $u_r$ are three vectors while $A$ is in  the adjoint of $SO(2n)$. The fields $M_{rs}$ and $v_r$ are singlets.
The term $\epsilon_{rst}(\dots)$ in the superpotential originates from the Pfaffian of the generalized meson, built up by contracting the fundamentals of the $USp(2n)$ gauge node, after integrating out the massive component $M_{Qa} = Q a$.

The partition function is obtained by the limiting case of the identity 
 given in {\bf Proposition 5.3.4} of \cite{VanDeBult}
 and we report it in formula (\ref{USp}). It corresponds to the confining duality for $USp(2n)$ with $2n+4$ fundamentals and linear monopole superpotential turned on.
This  identity was obtained also in \cite{Aharony:2013dha} from the reduction of the integral identity relating the superconformal indices of the 4d duality of \cite{Intriligator:1995ne}.
The partition function obtained after confining the $USp(2n)$ gauge
node is
\begin{eqnarray} \label{eq:SO(2n)_step2}
Z_{SO(2n)}
\!\!&=&\!\!
\frac{\prod_{r<s} \Gamma_h(\mu_r + \mu_s)
\prod_{r=1}^{3} \Gamma_h \left(\omega+n \tau- \frac{\tau}{2} + \mu_r \right)
\Gamma_h(2 \omega-n \tau)  \prod_{k=1}^{n-1} \Gamma_h(2 \omega-2k \tau)  }
{(-\omega_1 \omega_2)^\frac{n}{2} 2^{n-1} (n!)}
\nonumber \\
\!\!&\times&\!\!
\int 
\prod_{i=1}^{n} dx_i 
\prod_{r=1}^{3} \Gamma_h \left(\pm x_i + \mu_r + \frac{\tau}{2} \right) 
\prod_{1\leq i<j\leq n}  \frac{\Gamma_h(\pm x_i \pm x_j  + \tau)}{\Gamma_h(\pm x_i \pm x_j) }
\end{eqnarray}

As a consistency check we can now use formula (\ref{SOev}) on the
integral (\ref{eq:SO(2n)_step2})
because the mass parameters are constrained as in (\ref{BCSO}).
After some rearranging we eventually checked that 
the integral reduces to the LHS of (\ref{SPP}). This signals the consistency of the various steps done so far and motivated us to further deconfine the adjoint of $SO(2n)$ in order to produce a new quiver with a symplectic and an orthogonal  node.

\begin{figure}
\centering
\includegraphics[width=10cm]{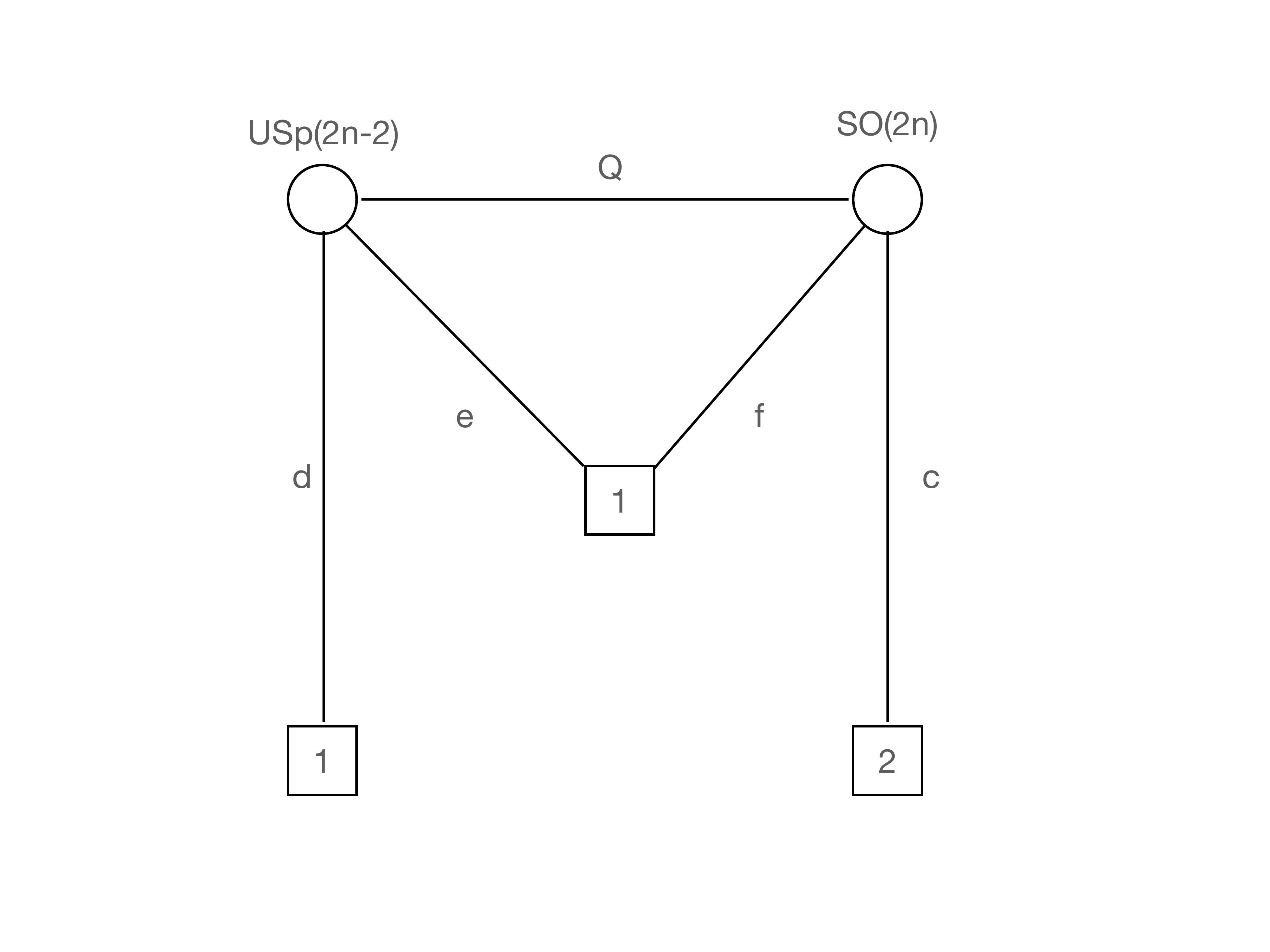}
\caption{Quiver representation of the $SO(2n)$ model after the adjoint field has been deconfined}
\label{fig:dec_quiver_2}
\end{figure}

The $SO(2n)$ model with adjoint and three fundamentals is equivalent
to the $USp(2n-2) \times SO(2n)$ quiver given in Figure \ref{fig:dec_quiver_2}
 with superpotential 
 \begin{eqnarray}
 W &=& Y_{USp}+ Y_{SO}^+ + 
 \sum_{k=1}^{n-1} \rho_k\Tr (Q Q)^{2k}
 +\Tr (Q e f)
 + v_1 \epsilon_{2n}(Q^{2n-2} c_2 c_3)
   \nonumber \\
 &+&
  v_3\Tr (f c_2 )+ v_2\Tr (f c_3) +\epsilon_{2n} (Q^{2n}) \, (Tr(d e)+ \epsilon_{rst}  M_{rs} v_t) 
   \nonumber \\
   \label{eq:W_T2prime}
 \end{eqnarray}
The duality map reflects on the following relations between the mass parameters in the partition function
 \begin{equation}
\mu_{c_{2,3}}  =  \mu_{2,3}+ \frac{\tau}{2},\quad
 \mu_{d_1} = \mu_{1},\quad
 \mu_Q = \frac{\tau}{2}
 \end{equation}
 Furthermore the superpotential imposes the following relations on the  other parameters
 \begin{equation}
 \mu_{f} = \omega-\mu_2-\mu_3-n\tau,\quad
 \mu_{e} = 2\omega-n \tau-\mu_1
 \end{equation}
 and the usual constraint
 \begin{equation}
 2n\tau-\frac{\tau}{2} + \sum_{r=1}^{3} \mu_a = \omega
 \end{equation}

 We can see that this model reduces to the $SO(2n)$ model discussed above when the $USp(2n-2)$ node with $2n+2$ fundamentals and a linear monopole superpotential  confines.
Again the confinement of the $USp(2n-2)$ symplectic gauge group gives raise to a superpotential term proportional to the Pfaffian of the generalized meson.
By integrating out the massive fields and substituting in the Pfaffian we recover the superpotential (\ref{SOSOSO}).
 The partition function of the $USp(2n-2) \times SO(2n)$ model is  
  \begin{eqnarray}
Z_{SO(2n) \times USp(2n-2)}
\!\!&=&\!\!
\frac{\prod_{r<s} \Gamma_h(\mu_r + \mu_s)
\prod_{r=1}^{3} \Gamma_h \left(\omega+n \tau- \frac{\tau}{2} + \mu_r \right)
\prod_{k=1}^{n-1} \Gamma_h(2\omega-k\tau) }
{(-\omega_1 \omega_2)^\frac{2n-1}{2} 2^{2n-2} (n!(n-1)!)}
\nonumber \\
\!\!&\times&\!\!
\int 
\prod_{i=1}^{n} dx_i 
\frac{\Gamma_h (\pm x_i + \mu_f) \prod_{\ell=2}^{3} \Gamma_h (\pm x_i + \mu_{c_\ell}) }{ \Gamma_h(\pm x_i)} 
\nonumber \\
\!\!&\times&\!\!
\prod_{a=1}^{n-1} dy_a  
 \frac{\Gamma_h (\pm y_a +\mu_{d})  \Gamma_h (\pm y_a +\mu_e) }{\Gamma_h(\pm 2 y_a)}
\nonumber \\
\!\!&\times&\!\!
\frac{
\prod_{i=1}^{n}\prod_{a=1}^{n-1} \Gamma_h \left(\pm x_i \pm y_a +\mu_Q \right)}
{\prod_{i<j} 
\Gamma_h(\pm x_i \pm x_j) \prod_{a<b} \Gamma_h(\pm y_a \pm y_b)}
\label{eq:Z_T2prime}
\end{eqnarray}

One can check that the partition functions for the $SO(2n)$ model and that for the $USp(2n-2)\times SO(2n)$ quiver are equal by applying the  identity for the confining $USp$ node with $2n+2$ fundamentals discussed above.
The last step consists in performing a confining duality on the $SO(2n)$ gauge node with $2n+1$ vectors and linear monopole superpotential turned on.
This gives raise to an $USp(2n-2)$ gauge theory with an adjoint, four fundamentals and a series of singlets.
The mesonic and baryonic operators associated to the $SO(2n)$ gauge group are 
\begin{eqnarray}
\label{mesbar}
\mathcal{M} = 
\left(
\begin{array}{ccc}
S\equiv Q^2 & M_{Q,f} \equiv Q f&M_{Q, c_l} \equiv Q c_l\\
M_{Q,f}^T &  M_{ff} \equiv f f &M_{f, c_l}\equiv f c_l \\
M_{Q, c_l}^T  & M_{f, c_l}^T&M_{ c_l,  c_m}\equiv c_l c_m
\end{array}
\right), \quad
\mathcal{Q} =
\left(
\begin{array}{c}
q \equiv \epsilon_{2n} (Q^{2n-3} f c_2 c_3) \\
q_f\equiv \epsilon_{2n} (Q^{2n-2}  c_2 c_3) \\
q_{c_2}\equiv\epsilon_{2n} (Q^{2n-2} f  c_3) \\
q_{c_3}\equiv\epsilon_{2n} (Q^{2n-2} f c_2 )
\end{array}
\right) \nonumber \\
\end{eqnarray}
with $l,m=2,3$ and
where $S$ is in the adjoint of the $USp(2n-2)$ gauge group, while $M_{Q,f}$, $M_{Q,c_1}$,  $M_{Q,c_2}$  and $q$  are four fundamentals of $USp(2n-2)$.
There are also two extra fundamentals of $USp(2n-2)$ corresponding to the fields $d$ and $e$ of the previous model, which are not modified by the duality on the $SO(2n)$ gauge node.
The superpotential of the dual  $USp(2n-2)$ adjoint SQCD  is then
\begin{eqnarray}
\label{final}
W &=& Y_{USp}+ \sum_{k=1}^{n-1} \rho_k\Tr S^{2k} +  \mathcal{M} \mathcal{Q}  \mathcal{Q}^T  + \det \mathcal{M}
+ e M_{Q,f} + v_1 q_f \nonumber \\
&+&
 v_3 M_{f,c_2} + v_2 M_{f,c_3} +\Tr\left(S^n \right)  (d e + 
\epsilon_{rst} M_{rs} v_t)  
\end{eqnarray}
The determinant can be evaluated as
\begin{eqnarray}
W_{\det \mathcal{M}} &=& \det S \det 
\left(
\begin{array}{cc}
M_{ff}  &M_{f,\vec c} \\
M_{f,\vec c}^T&M_{\vec c, \vec c}
\end{array}
\right)
+\epsilon_{2n-2} \epsilon_{2n-2} \left(S^{2n-3}(
M_{Q,f} M_{c_2,c_3} + M_{Q c_2} M_{f,c_3} 
\right. \nonumber \\
&+&
 M_{Q c_3} M_{f,c_2})^2
 +\left.
S^{2n-4} (M_{Q,f} M_{Q,c_2} M_{Q,c_3} 
(
M_{Q,f} M_{c_2,c_3} + M_{Q c_2} M_{f,c_3} 
\right. \nonumber \\
&+&
M_{Q c_3} M_{f,c_2})
+\left.
S^{2n-5} (M_{Q,f} M_{Q,c_2} M_{Q,c_3})^2
\right)
\end{eqnarray} 
We can then integrate out the massive fields
$\{ e, M_{Q,f}, \vec v , q _f,M_{f,\vec c} \} $ and we are left with $USp(2n-2)$ 
adjoint  SQCD with four fundamentals.
There is a rather rich structure of singlets that we do not report here but that can be read by computing the $F$-terms of (\ref{final}).
We can now iterate this procedure by alternating adjoint deconfinement and 
duality in order to arrive to the final step and eventually prove 
the duality. 

As anticipated this procedure can be used on the mathematical side to prove the identity (\ref{SPP}) from a physical perspective. In order to complete the proof we need to consider the partition function obtained so far after the final duality on the $SO(2n)$ node \eqref{SOeven}. It is
\begin{equation}
\begin{split}
	& Z_{USp(2n-2)} = 
	\frac{
		\prod_{l,m=1,2} \Gamma_h(\mu_{c_l}+\mu_{c_m})
		\prod_{l=1}^2 \Gamma_h \left(\omega-\mu_{c_l}\right)
		\Gamma_h (2\mu_a)
		\prod_{r<s}^{3} \Gamma_h \left(\mu_r+\mu_s\right) }
		{(-\omega_1 \omega_2)^{\frac{n-1}{2}}  2^{n-1} (n-1)!}
	\\
	\times&
	\prod_{k=1}^{n-1} \Gamma_h(2 \omega-2 k \tau)
	\int 
	\prod_{\alpha=1}^{n-1} dy_\alpha 
	 \frac{\prod_{r=1}^4 \Gamma_h (\pm y_\alpha +{\tilde \mu_r})   }{\Gamma_h(\pm 2 y_\alpha)}
	 \prod_{1\leq \alpha<\beta\leq n-1}\frac{\Gamma_h \left(\pm y_\alpha \pm y_\beta + \tau \right)}
	 					{\Gamma_h \left(\pm y_\alpha \pm y_\beta  \right)}
\end{split}
\label{eq:Z_final}
\end{equation}
Where the masses ${\tilde \mu}_r$ are:
\begin{equation}	\label{eq:mass_redef}
	\vec{\tilde{\mu}} = \left\{ \mu_1, \, \mu_2 + \tau, \, \mu_3 + \tau, \, \omega - \frac{\tau}{2} \right\}
\end{equation}
Notice that the superpotential constraint reads:
\begin{equation}
	2(n-1) \tau+ \sum_{r=1}^4 \tilde{\mu}_r =2 \omega \quad
	\& \quad
	2 \tilde \mu_4+\tau=2\omega
\end{equation}
which is equivalent in form to the original superpotential constraint 
\eqref{original}.
\\
The contribution of the singlets can be written as:
\begin{equation}	\label{eq:singlets_USp}
\begin{split}
	&\prod_{k=1}^{n-1} \Gamma_h \left( 2\omega-2 k\tau \right)
	\prod_{r<s}^{3} \Gamma_h \left(\mu_r+\mu_s\right) 
	\prod_{r=2}^3
	\Gamma_h \left( \mu_{r} + \mu_1 (2n-1) \tau \right)
	\\
	&\times 
	\Gamma_h \left( \mu_2 + \mu_3 + \tau \right)
	\prod_{r=2}^3 \Gamma_h \left( 2 \mu_r + \tau \right)
	\Gamma_h \left( 2 \mu_1 + \tau (2n-1) \right)
\end{split}
\end{equation}

We can prove the confining duality for $USp(2n)$ with four fundamental and linear monopole superpotential by iterating this procedure $n$ times. In each step we obtain a new set of singlets as in \eqref{eq:singlets_USp}, with the exception that the tower of $\Gamma_h \left( 2\omega-2 k\tau \right)$ reduces of one unit. Furthermore in each step the rank of the gauge group decreases by one and the real masses are redefined as in \eqref{eq:mass_redef}, so that the fundamentals of $USp(2(n-h))$ obtained after $h$ steps are related to the original ones by:
\begin{equation}	\label{eq:mass_redef2}
	\vec{\mu}_{\text{$h$-th step}} = \left\{ \mu_1, \, \mu_2 + h \tau, \, \mu_3 + h \tau, \, \omega - \frac{\tau}{2} \right\}
\end{equation}

Thus iterating this procedure $n$ times each term in \eqref{eq:singlets_USp} gives a tower of singlets of the final confined phase. Schematically:

\begin{eqnarray}
\prod_{r<s}^{3} \Gamma_h \left(\mu_r+\mu_s\right)  \quad& \to &\quad \left \lbrace
	\begin{split}
		&\prod_{\ell=0}^{n-1}\prod_{r=2}^3 \Gamma_h \left( \ell \tau + \mu_1 + \mu_r \right)
		\\
		&\prod_{\ell = 0}^{n-1} \Gamma_h \left(2\ell\tau + \mu_2 + \mu_3 \right)
	\end{split} \right.
\\
\prod_{r=2}^3 \Gamma_h \left( \mu_{r} + \mu_1 (2n-1) \tau \right) \quad& \to &\quad 
\prod_{\ell=n}^{2n-1} \prod_{r=2}^3 \Gamma_h \left( \ell\tau + \mu_1 + \mu_r \right)
\\
\Gamma_h \left( \mu_2 + \mu_3 + \tau \right)\quad& \to &\quad 
\prod_{\ell =0}^{n-1} \Gamma_h \left( (2\ell+1) \tau + \mu_2 + \mu_3 \right)
\\
\prod_{r=2}^3 \Gamma_h \left( 2 \mu_r + \tau \right)\quad& \to &\quad 
\prod_{\ell=0}^{n-1}\prod_{r=2}^3  \Gamma_h \left( 2 \mu_r + (2\ell+1)\tau \right)
\\
\Gamma_h \left( 2 \mu_1 + \tau (2n-1) \right)\quad& \to &\quad 
\prod_{\ell=0}^{n-1}  \Gamma_h \left( 2 \mu_1 + \tau (2\ell+1) \right)
\end{eqnarray}
while the contribution of the tower $\prod_{k=1}^{n-1} \Gamma_h \left( 2\omega-2 k\tau \right)$ reduces of one unit at each step, and eventually disappear.
Together these reproduce the formula \eqref{SPP}.
%
%
\subsubsection*{Case II: $SO(2n)$}
%
%
Now we prove the confining duality for SO$(2n)$ with one adjoint $A$, three vectors $q_{1,2,3}$ and monopole superpotential \eqref{SOev} by deconfining the adjoint. The mass parameters  for the three vectors $q_r$  are referred as $\mu_r$ with $r=1,2,3$ and the one for the adjoint is referred as $\tau$.
The $SO(2n)$ model is equivalent to the $USp(2n-2)\times SO(2n)$ quiver in Figure \ref{fig:dec_quiver_2}, but this time the superpotential is
\begin{equation}
\label{quivT2}
W = Y_{USp} + Y_{SO}^+ + g\Tr (de) +\Tr (Qef)
\end{equation}
The duality map is:
\begin{equation}
\mu_{c_{2,3}} = \mu_{2,3},
\quad
\mu_d = \mu_1 - \frac{\tau}{2},
\quad
\mu_Q = \frac{\tau}{2}
\end{equation}

The other parameters are fixed by the constraints given by the superpotential:
\begin{equation}
\mu_e = 2 \omega -n \tau -\mu_1 + \frac{\tau}{2}
	\quad
	\mu_f = \omega - \mu_2 - \mu_3 - (n-1)\tau = (n-1)\tau + 	\mu_1,
	\quad 
	\mu_g = n \tau
\end{equation}
with the constraint given by the monopole superpotential:
\begin{equation}
\label{bcfin}
	2(n-1)\tau + \sum_{r=1}^{3} \mu_r = \omega
\end{equation} 
The partition function of the quiver is:
\begin{eqnarray}
Z_{SO(2n) \times USp(2n-2)}
\!\!&=&\!\!
\frac{ \Gamma_h \left( \mu_g \right) }
{(-\omega_1 \omega_2)^\frac{2n-1}{2} 2^{2n-2} (n!(n-1)!)}
\nonumber \\
\!\!&\times&\!\!
\int 
\prod_{i=1}^{n} dx_i 
\frac{\Gamma_h (\pm x_i + \mu_f) \prod_{\ell=2}^{3} \Gamma_h (\pm x_i + \mu_{c_\ell}) }{ \Gamma_h(\pm x_i)} 
\nonumber \\
\!\!&\times&\!\!
\prod_{a=1}^{n-1} dy_a  
 \frac{\Gamma_h (\pm y_a +\mu_d)  \Gamma_h (\pm y_a +\mu_e) }{\Gamma_h(\pm 2 y_a)}
\nonumber \\
\!\!&\times&\!\!
\frac{
\prod_{i=1}^{n}\prod_{a=1}^{n-1} \Gamma_h \left(\pm x_i \pm y_a +\mu_Q \right)}
{\prod_{i<j} 
\Gamma_h(\pm x_i \pm x_j) \prod_{a<b} \Gamma_h(\pm y_a \pm y_b)}
\end{eqnarray}

Now we dualize the node with orthogonal group, this results in a $USp(2n-2)$ model with four fundamentals and  superpotential:
\begin{equation}
W = Y_{USp} + h\Tr (de) +\Tr (M_{Qf} e)
+\det \mathcal{M} +Tr \mathcal{Q} \mathcal{M} \mathcal{Q} 
\end{equation}
where $\mathcal{M}$ and $\mathcal{Q}$ are given by
(\ref{mesbar}).
Due to the rather complicated structure of such superpotential 
we decide to proceed by adding some interactions in the original theory.
We turn on the extra superpotential term 
\begin{equation}
\label{defso}
\delta W_{SO(2n)} = 
\sum_{k=1}^{n-1} \rho_k \Tr A^{2i} + 
\beta \text{ Pf } A+ \epsilon_{rst}  \alpha_r \epsilon_{2n}(A^{n-1} q_s q_t)
\end{equation}
On the partition function this removes the contributions
of $\Gamma_h(n\tau)$,  $\prod_{r<s} \Gamma_h((n-1)\tau+\mu_r+\mu_s)$  and $\prod_{k=1}^{n} \Gamma_h(2k \tau)$ from the RHS of (\ref{SOev})
giving raise to the contributions $ \Gamma_h(2\omega-n \tau)$,  $\prod_{r=1}^3 \Gamma_h(\omega+(n-1)\tau+\mu_r)$ 
and $\prod_{k=1}^{n} \Gamma_h(2\omega-2k \tau)$  
on the LHS.
Mathematically this is achieved by applying the reflection
equation and the balancing condition (\ref{bcfin}) and it does not spoil the integral identity  (\ref{SOev}).
Furthermore  (\ref{quivT2}) becomes
\begin{eqnarray}
\label{quivT22}
W &=& Y_{USp} + Y_{SO}^+ + \sum_{k=1}^{n-1} \rho_k\Tr Q^{2k}  +\Tr (Qef)
\nonumber \\
&+&
\alpha_1 \epsilon_{2n} (Q^{2n-2} c_2 c_3)+ \alpha_2 \Tr (f c_2) + \alpha_3 \Tr (f c_3)
\end{eqnarray}
In this way we can dualize the $USp(2n-2)$ node integrating out  $M_{Qe}$
and $f$ and identify $\beta$ with $M_{de}$. The final result coincides to the
original model with the superpotential deformation (\ref{defso}).

We can proceed by confining the $SO(2n)$ node with $2n+1$ fundamentals and linear monopole superpotential after we have added the contributions of 
$\alpha_{1,2,3}$ and $\beta$.
The partition function for the $USp(2n-2)$ model
is
\begin{equation}
\begin{split}
	Z_{USp(2n-2)}
	\!\!&=\!\!
	\frac{\prod_{k=1}^{n-1} \Gamma_h(2\omega- 2 k \tau)
		\prod_{2 \leq l\leq m \leq 3} \Gamma_h(\mu_{c_l}+\mu_{c_m})
		\prod_{l=2}^3 \Gamma_h \left(\omega-\mu_{c_l}\right)
		\Gamma_h (2\mu_f) }
		{(-\omega_1 \omega_2)^{\frac{n-1}{2}}  2^{n-1} (n-1)!}
	\\
	&\times
	\int 
	\prod_{a=1}^{n-1} dy_a  
	 \frac{\prod_{r=1}^4 \Gamma_h (\pm y_a +{\tilde \mu}_r)   }{\Gamma_h(\pm 2 y_a)}
	 \prod_{1\leq a<b\leq n-1}\frac{\Gamma_h \left(\pm y_a \pm y_b + \tau \right)}
	 					{\Gamma_h \left(\pm y_a \pm y_b  \right)}
\end{split}
\end{equation}
Where the masses are:
\begin{equation}
	\vec{\tilde{\mu}} = \left\{ \mu_1 - \frac{\tau}{2}, \, \mu_2 + \frac{\tau}{2}, \, \mu_3 + \frac{\tau}{2},\, \omega- \frac{\tau}{2}\right\}
\end{equation}

If we now ignore the singlets we observe that the contribution of the
 $USp(2n-2)$ gauge sector to this partition function
 corresponds to the LHS of the identity \eqref{SPP}. The duality associated to such a sector  was proven in the previous section. We can then use this duality to confine the $USp(2n-2)$ theory, resulting in a WZ model with partition function:
\begin{equation}
\begin{split}
	&
	\prod_{2 \leq l\leq m \leq 3} \Gamma_h(\mu_{c_l}+\mu_{c_m})
		\cdot \Gamma_h (2\mu_f) \cdot
	\prod_{\ell=0}^{n-2} \prod_{r=1}^{2} \Gamma_h\left( 2(\ell+1)\tau +2 { \mu}_{r} \right)
		 \Gamma_h\left( 2 \ell\tau +2 { \mu}_1 \right)
	\\
	&\times
	\prod_{l=2}^3 \Gamma_h \left(\omega-\mu_{c_l}\right)	\cdot
	\prod_{\ell=0}^{2n-3} \big(\Gamma_h\left( (\ell+1)\tau + { \mu}_2 + { \mu}_3 \right) \cdot
		 \prod_{r=1}^{2}  \Gamma_h\left( \ell\tau + { \mu}_1 + { \mu}_{r}\right)
		 \big)
\end{split}
\end{equation}
which reproduces the RHS of \eqref{SOev} once the contributions of the baryons $\text{Pf}\, A$ and $\epsilon_{2n-2} (A^{n-1} q_r q_s)$ and of the singlets $\Tr A^{2k}$ are removed. 
%
%
\subsubsection*{Case III: $SO(2n+1)$}
%
%

\begin{figure}
\centering
\includegraphics[width=10cm]{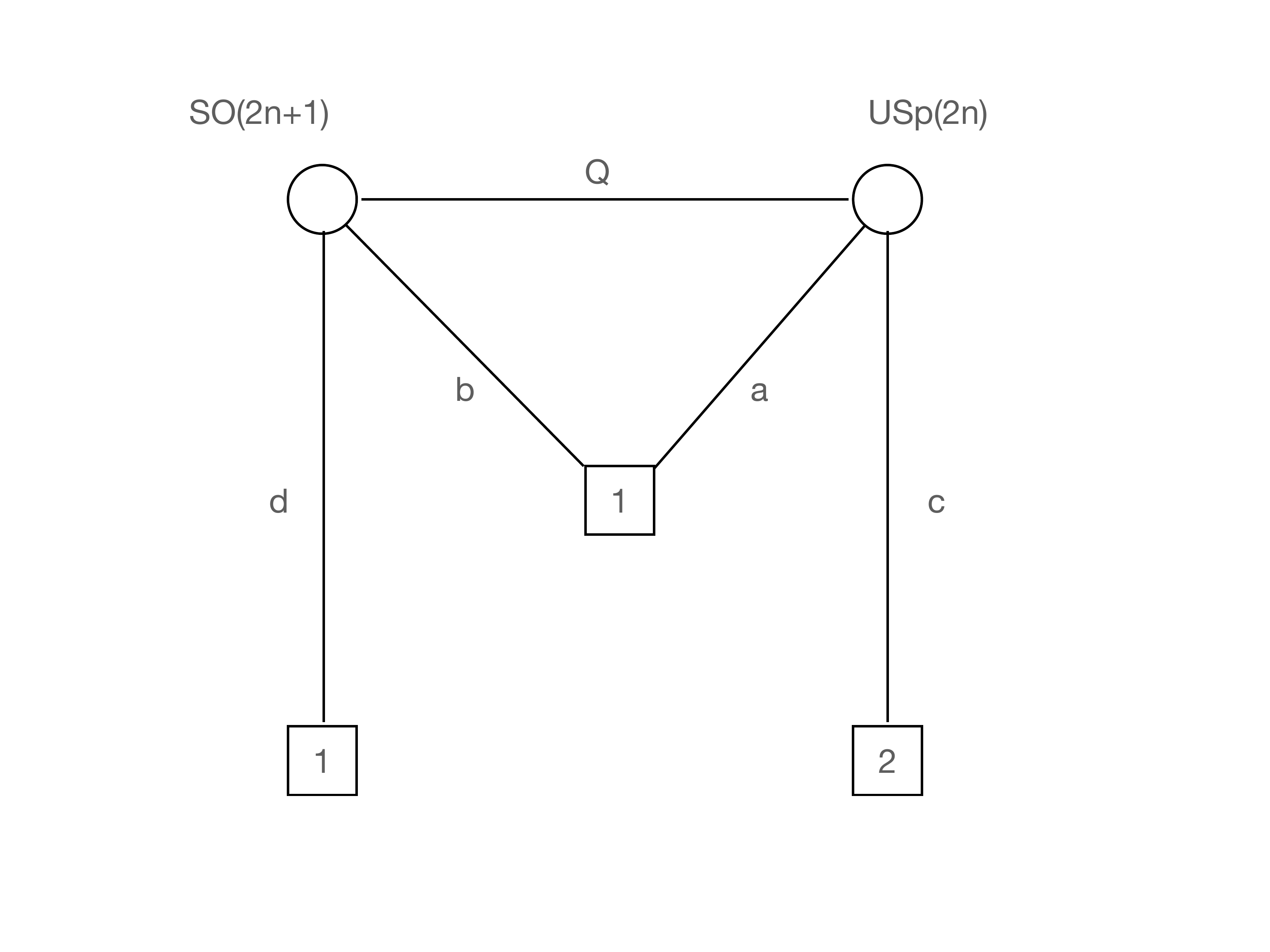}
\caption{Quiver representation of the $SO(2n+1)$ model after the adjoint field has been deconfined}
\label{fig:dec_quiver_3}
\end{figure}

The $SO(2n+1)$ model with adjoint $A$ and three fundamentals $q_{1,2,3}$ is equivalent
to the $USp(2n) \times SO(2n+1)$ quiver given in Figure  \ref{fig:dec_quiver_3}.
The superpotential of this quiver is given by
\begin{equation}
\label{lastquiv}
W = Y_{USp} + Y_{SO}^++Tr(a b Q) + f_1 Tr(c_1 a) + f_2 Tr(c_2 a)+ g Tr(c_2 c_3)
\end{equation}
The mass parameters in the partition function are 
\begin{eqnarray}
&&
\mu_{c_{2,3}} = \mu_{2,3} -\frac{\tau}{2} ,\quad
~~~\mu_Q = \frac{\tau}{2},\quad \mu_d = \mu_1,\quad
~~~~~\mu_a =  2 \omega-\mu_2-\mu_3-n \tau+\frac{\tau}{2} 
\nonumber \\
&&
\mu_b  = \omega-\mu_1-n \tau ,\quad
\mu_g  = 2\omega-\mu_2-\mu_3-\tau,\quad \mu_{f_{2,3}} = \mu_{3,2}+ n\tau  \nonumber \\
\end{eqnarray}
with the constraint (\ref{constrodd}).
The partition function of the $SO(2n+1) \times USp(2n)$ quiver is given by
 \begin{eqnarray}
Z_{SO(2n+1) \times USp(2n)}
\!\!&=&\!\!
\frac{  \Gamma_h(\mu_{g}) \prod_{m=2,3} \Gamma_h(\mu_{f_m})) 
\Gamma_h ( \mu_d ) \Gamma_h ( \mu_b )}
{(-\omega_1 \omega_2)^\frac{2n}{2} 2^{2n} (n!)^2)}
\nonumber \\
\!\!&\times&\!\!
\int 
\prod_{i=1}^{n} dx_i 
\frac{\Gamma_h (\pm x_i + \mu_d) \Gamma_h (\pm x_i + \mu_b)    }{ \Gamma_h(\pm x_i)} 
\nonumber \\
\!\!&\times&\!\!
\prod_{\alpha =1}^{n} dy_\alpha  
 \frac{\Gamma_h (\pm y_\alpha +\mu_a) \prod_{m=2}^{3}  
 \Gamma_h (\pm y_\alpha +\mu_{c_m}) \Gamma_h (\pm y_\alpha + \mu_Q)}{\Gamma_h(\pm 2 y_\alpha)}
\nonumber \\
\!\!&\times&\!\!
\frac{
\prod_{i=1}^{n}\prod_{\alpha=1}^{n} \Gamma_h \left(\pm x_i \pm y_\alpha +\mu_Q \right)}
{\prod_{i<j} 
\Gamma_h(\pm x_i \pm x_j) \prod_{\alpha<\beta} \Gamma_h(\pm y_\alpha \pm y_\beta)}
\end{eqnarray}
Next we have to confine the $SO(2n+1)$ sector with $2n+2$ vectors and a linear monopole superpotential  and we end up with $USp(2n)$.
The problem consists of understanding the interaction among the various singlets from the confining dynamics of $SO(2n+1)$.
Again we can simplify the problem by modifying the original $SO(2n+1)$ model by considering the superpotential
\begin{equation}
W = Y_{SO}^+ + \sum_{k=1}^{n} \rho_k \Tr A^{2k}+ \beta \epsilon_{2n+1} (A^{n-1} q_1 q_2 q_3) + \sum_{r=1}^{3} \alpha_r \epsilon_{2n+1} (A^{n} q_r)
\end{equation}
corresponding to remove the baryons and the singlets $\Tr A^{2k}$ from the confined phase and add the new singlets $\alpha_{1,2,3}$ and $\beta$ in the original model.
On the partition function this removes the contributions
of $\Gamma_h(\omega- n\tau)$, $\prod_{r=1}^{3} \Gamma_h(n\tau+\mu_r)$   and $\prod_{k=1}^{n} \Gamma_h(2k \tau)$ from the RHS of (\ref{SOddd})
giving raise to the contributions $ \Gamma_h(\omega+n \tau)$,
$\prod_{r=1}^{3} \Gamma_h(2 \omega-n\tau-\mu_r)$
and $\prod_{k=1}^{n} \Gamma_h(2\omega-2k \tau)$  
on the LHS.
Mathematically this is achieved by applying the reflection
equation and it does not spoil the integral identity  (\ref{SOddd}).
By deconfining the adjoint $A$ the superpotential (\ref{lastquiv}) is modified as well.
The new superpotential is 
\begin{equation}
\label{lastquiv2}
W = Y_{USp} + Y_{SO}^++Tr(a b Q) + g Tr(c_2 c_3) + \beta\Tr b d
+ \alpha_1 \epsilon_{2n+1} (Q^{2n} d)
\end{equation}
We can proceed by confining the $SO(2n+1)$ node. By integrating out the massive fields we arrive to 
an $USp(2n)$ gauge theory with an adjoint $S$, three fundamentals,
identified by $d$ and the two mesonic composites $Q c_{2}$ and $Q c_{3}$,
and a fourth fundamental corresponding to $u=\epsilon_{2n+1} (Q^{2n-1} b d)$ , interacting with the adjoint through a superpotential term $W\propto u S u$.
There is also a linear monopole superpotential and many more interactions with the singlets that we do not report here, but that can obtained by evaluating the determinant $\det S$ and the superpotential contraction 
of $S$ with the baryons of the confined $SO(2n+1)$ node.
The partition function of the model is 
 \begin{eqnarray}
 \label{USPlast}
Z_{USp(2n)}
\!\!&=&\!\!
\frac{ \Gamma_h(\tau)^n \prod_{k=1}^{n} \Gamma_h(2\omega-2 k\tau)  
\Gamma_h(2 \mu_1)   \Gamma_h(\omega- \mu_1)\Gamma_h(\omega+\mu_1+2n \tau)   
  \Gamma_h(2\mu_b)  }
{(-\omega_1 \omega_2)^\frac{n}{2} 2^{n} (n!))}
\nonumber \\
\!\!&\times&\!\!
\int
\prod_{a=1}^{n} dy_a  
\frac{ \prod_{r=1}^4\Gamma_h (\pm y_i +\tilde \mu_r) }{\Gamma_h(\pm 2 y_a)}
\prod_{1\leq a<b\leq n}\frac{ \Gamma_h(\pm y_a \pm y_b+\tau )}{ \Gamma_h(\pm y_a \pm y_b)}
\end{eqnarray}
with $\vec{\tilde \mu} = \left\{\mu_1-\frac{\tau}{2},\mu_2+\frac{\tau}{2},\mu_3+\frac{\tau}{2},\omega-\frac{\tau}{2} \right\}$
and the constraints
 $\sum_{\ell=1}^4 \tilde \mu_\ell +2 n \tau = 2 \omega$ 
 and $2 \tilde \mu_4 +\tau=2 \omega$.
Also in this case we can borrow the results of the previous sections.
Indeed if we  ignore the singlets we observe that the contribution of the
gauge $USp(2n-2)$ sector to this partition function
 corresponds to the LHS of the identity \eqref{SPP}. The duality associated to such a sector  was proven in the previous section. We can then use this duality to confine the $USp(2n-2)$ theory and prove the confining duality for the $SO(2n+1)$ model.

\section{Conclusions}
\label{sec:conc}

In this paper we have studied 3d $\mathcal{N}=2$ confining 
gauge theories with real USp/SO gauge groups, with 
fundamentals/vectors and adjoint matter.
We have first shown that the symplectic and orthogonal cases recently studied in \cite{Benvenuti:2021nwt}, with two fundamentals and one vector
respectively, can be studied by the squashed three-sphere localization
by applying the \emph{duplication formula} for the hyperbolic Gamma function of another s-confining model, namely $USp(2n)$ with an antisymmetric and four fundamentals.
 Motivated by this relation we then elaborated on the case of 
$USp(2n)$ with an antisymmetric, six fundamentals and a monopole superpotential.
By applying the same strategy we derived three new integral identities 
involving 
symplectic and orthogonal adjoint SQCD, with four fundamentals and three vector respectively and a monopole superpotential.
We showed that the new confining cases reduce to the ones of 
\cite{Benvenuti:2021nwt} by a real mass flow and then we proved the dualities by sequentially deconfining the adjoint fields.
This last step furnished an alternative proof of the identities
and (\ref{SPP}), (\ref{SOev})  and (\ref{SOddd}) as we have explicitly shown.

This paper is the starting point of many further analysis.

For example one can apply the \emph{duplication formula} to the  integral identities for $USp(2n)$ theories with
an antisymmetric and eight fundamentals, where the $A_7$ global symmetry enhances to $E_7$. 
This case has been deeply investigated in the mathematical \cite{VanDeBult}
and then in the physical literature \cite{Amariti:2018wht,Benvenuti:2018bav} and it may be interesting to understand if similar enhancements or new dualities appear for 
models with adjoint matter as well.

Another interesting family of models that may deserve some further investigation are models with power law superpotential for the two index tensor. In this case the starting point of the analysis are the integral identities discussed in \cite{Amariti:2015vwa} for $USp(2n)$ with antisymmetric and adjoint matter fields. Again applying the duplication formula in such cases could lead to new  relations between these models and to new results for the orthogonal cases.

A deeper question that we have not addressed here consists of  the physical
interpretation, if any, of the \emph{duplication formula}.
As observed in the literature this formula allows to switch from the integral identities for the $USp(2n)$ duality with fundamentals to the integral identities for the $SO(n)$ dualities with vectors. This has been discussed in \cite{Spiridonov:2011hf} for the superconformal index of 4d dualities and in  \cite{Benini:2011mf} for the squashed three-sphere partition  function of 3d dualities.
In presence of monopole superpotential this issue is more delicate, because 
in some cases it can lead to a singular behavior that requires more care.
In any case, when the procedure gives rise to a finite result, also in presence of monopole superpotential, the constraints imposed by anomalies  (in 4d) and by monopole superpotential (in 3d) translate in a consistent way into the new identities, and the latter can be interpreted as new physical dualities (or in new examples of s-confining theories).
It should be then important to have a physical interpretation of the 
\emph{duplication formula}.

A last comment is related to the adjoint deconfinement and to a possible relation with another mathematical result, that  consists of interpreting the various steps discussed when deconfining the adjoints as a manifestation of a generation of a chain or a tree of identities, along the lines of the Bailey's lemma. Such analysis has been first applied to the study of elliptic hypergeometric integrals (i.e. to the 4d superconformal index) in \cite{spiridonov2004inversions,2004}.
Recently a 4d physical interpretation of such mechanism has been discussed in \cite{Brunner:2017lhb}. It should be interesting to develop similar results in our 3d setup for the deconfinement of the adjoints in the hyperbolic hypergeometric integrals.

%
%
%
%
%
%
\section*{Acknowledgments}
%
%
We are grateful to Sergio Benvenuti for comments on the manuscript.
This work has been supported in part by the Italian Ministero dell'Istruzione, 
Universit\`a e Ricerca (MIUR), in part by Istituto Nazionale di Fisica Nucleare (INFN) through the “Gauge Theories, Strings, Supergravity” (GSS) research project and in
part by MIUR-PRIN contract 2017CC72MK-003.

\appendix
%
%
\section{Dualities with adjoint and without $W_{\text{monopole}}$  on $Z_{S^3}$}
\label{sec:A}
%
%

Here we follow the sequential deconfinement procedure performed in Section 5.1, 5.2 and 5.3 of  \cite{Benvenuti:2021nwt} on the partition function. These chains of confining/deconfining dualities allows to prove the dualities for symplectic (orthogonal) gauge group with two fundamentals (one vector), one adjoint without monopole superpotential. The identities needed are 
\begin{equation}	\label{eq:aharony_SO_odd_2}
	\begin{split}
		Z_{SO(2n+1)}^{N_f=2n}&=
		\frac{\prod_{r=1}^{2n} \Gamma_{h}\left(\mu_r \right)}
		{\sqrt{-\omega_{1} \omega_{2}}^{n} 2^{n} n !}  
		\int_{C^{n}} \frac{\prod_{j=1}^{n} \prod_{r=0}^{2n} \Gamma_{h}\left(\mu_{r} \pm x_{j}\right)}
		{\prod_{1 \leq j<k \leq n} \Gamma_{h}\left(\pm x_{j} \pm x_{k}\right)  \prod_{i=1}^n \Gamma_{h}\left(\pm x_i\right) } 
		\prod_{j=1}^{n} d x_{j}
		\\
		&=
		\Gamma_{h}\left( \omega-\sum_{r=1}^{2n} \mu_{r}\right) 
		\prod_{1 \leq r\leq s \leq 2n} \Gamma_{h}\left(\mu_{r}+\mu_{s}\right)
		\prod_{r=1}^{2n} \Gamma_{h}\left(\omega - \mu_r\right)
	\end{split}
\end{equation}

\begin{equation}	\label{eq:aharony_USp}
	\begin{split}
		Z_{USp(2n)}^{N_f=2n+2}&=
		\frac{1}{\sqrt{-\omega_{1} \omega_{2}}^{n} 2^{n} n !} 
		\int_{C^{n}} \frac{\prod_{a=1}^{n} \prod_{r=0}^{2n+2} \Gamma_{h}\left(\mu_{r} \pm y_{a}\right)}{\prod_{1 \leq a<b \leq n} \Gamma_{h}\left(\pm y_{a} \pm y_{b}\right) \prod_{a=1}^{n} \Gamma_{h}\left(\pm 2 y_{a}\right)} \prod_{a=1}^{n} d y_{a}
		\\
		&=
		\Gamma_{h}\left( 2\omega-\sum_{r=1}^{2n+2} \mu_{r}\right) \prod_{1 \leq r<s \leq 2n+2} \Gamma_{h}\left(\mu_{r}+\mu_{s}\right)
	\end{split}
\end{equation}

\begin{equation}	\label{eq:aharony_SO_even_2}
	\begin{split}
		Z_{SO(2n)}^{N_f=2n-1} &=
		\frac{1}{\sqrt{-\omega_{1} \omega_{2}}^{n} 2^{n-1} n !} 
		\int_{C^{n}} \frac{\prod_{j=1}^{n} \prod_{r=0}^{2n-1} \Gamma_{h}\left(\mu_{r} \pm x_{j}\right)}
		{\prod_{1 \leq j<k \leq n} \Gamma_{h}\left(\pm x_{j} \pm x_{k}\right) } \prod_{j=1}^{n} d x_{j}
		\\
		&=
		\Gamma_{h}\left( \omega-\sum_{r=1}^{2n-1} \mu_{r}\right) 
		\prod_{1 \leq r\leq s \leq 2n-1} \Gamma_{h}\left(\mu_{r}+\mu_{s}\right)
		\prod_{r=1}^{2n-1} \Gamma_{h}\left(\omega-\mu_r \right)
	\end{split}
\end{equation}
which correspond to limiting cases of Aharony duality.

\subsubsection*{Case I: $USp(2n)$}
\begin{figure}
\centering
\includegraphics[width=10cm]{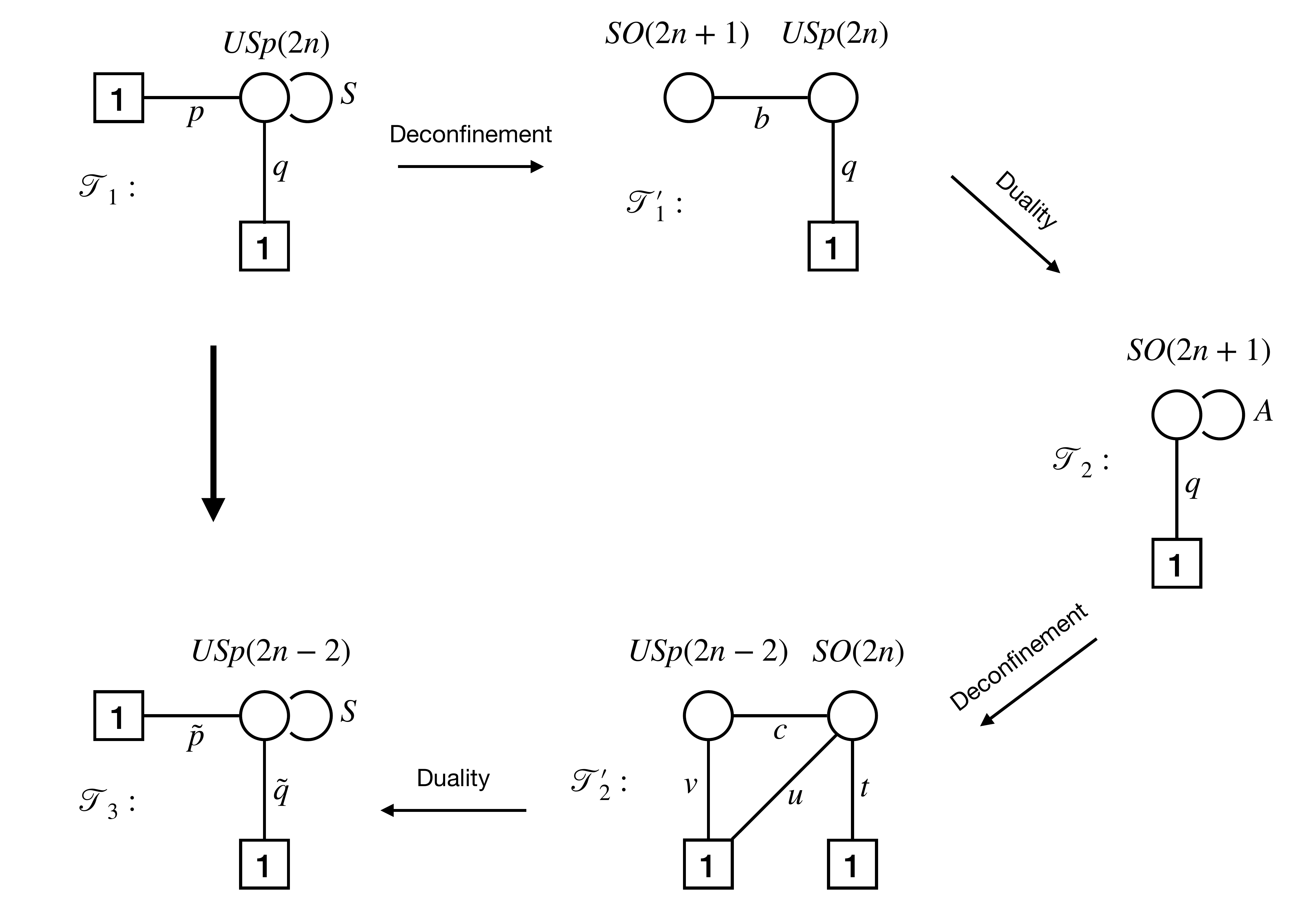}
\caption{Schematic representation of one step of the deconfinement procedure of \cite{Benvenuti:2021nwt} for the $USp(2n)$ model with adjoint. The partition functions of the model are:}
\vspace{10pt}
\begin{tabular}{||c||c|c|c|c|c||}
\hline
	&$ \mathcal{T}_1$	&	$ \mathcal{T}_1	'$	&$ \mathcal{T}_2$	&$ \mathcal{T}_2'$	&$ \mathcal{T}_3$ \\ \hline
$\mathcal{Z}_{S^3}$ & \eqref{eq:BLM_T1} &\eqref{eq:BLM_T1prime} &\eqref{eq:T2_BLM} & \eqref{eq:BLM_T2prime}&\eqref{eq:T3_BLM}\\
\hline
\end{tabular}
\label{fig:dec_quiver_4}
\end{figure}

	The partition function of theory $\mathcal{T}_1$ of \cite{Benvenuti:2021nwt} is:\\
	\tikzmark{T1}
	\begin{equation}
		\begin{split}
			Z_{\mathcal{T}_1}
			=&
			\frac{ \Gamma_h(\tau)^n }
			{(-\omega_1 \omega_2)^\frac{n}{2} 2^{n} n!}
			\int
			\prod_{a=1}^{n} dy_a  
			\frac{ \Gamma_h (\pm y_a+ m) \Gamma_h\left(\pm y_a + \omega - \frac{\tau}{2}\right) }
			{\Gamma_h(\pm 2 y_a)}
			\\&\times
			\prod_{1\leq a<b \leq n}\frac{ \Gamma_h(\pm y_a \pm y_b+\tau )}{ \Gamma_h(\pm y_a \pm y_b)} 
		\end{split}
		\label{eq:BLM_T1}
	\end{equation}
	This is equivalent to a two-node quiver with gauge groups $SO(2n+1)\times USp(2n)$, denoted $\mathcal{T}_{1'}$ with partition function:\\	
	\tikzmark{T1'}
	\begin{equation}
		\begin{split}
			Z_{\mathcal{T}_{1'}}
			=&
			\frac{ \Gamma_h(\tau)^n \Gamma_h (\omega + n\tau)}
			{(-\omega_1 \omega_2)^n 2^{2n} n!^2}
			\int
			\prod_{a=1}^{n} dy_a  
			\frac{ \Gamma_h (\pm y_a+ m) \Gamma_h \left( \pm y_a +\frac{\tau}{2}\right) }
			{\Gamma_h(\pm 2 y_a)}
			\prod_{i=1}^{n} dx_i
			\frac{ \Gamma_h\left( \pm y_a \pm x_i + \frac{\tau}{2}\right)}
			{\Gamma_h(\pm x_i)}
			\\&\times
			\prod_{1\leq a<b \leq n}\frac{1}{ \Gamma_h(\pm y_a \pm y_b)} 
			\prod_{1\leq i<j \leq n}\frac{1}{ \Gamma_h(\pm x_i \pm x_j)} 
		\end{split}
	\label{eq:BLM_T1prime}
	\end{equation}	
	These two expressions can be shown to coincide by using \eqref{eq:aharony_SO_odd_2} to confine the orthogonal node. Then we dualize the symplectic node using \eqref{eq:aharony_USp}:\\
	\tikzmark{T2}
	\begin{equation} 	\label{eq:T2_BLM}
		\begin{split}
			Z_{\mathcal{T}_2}
			=&
			\frac{ \Gamma_h(\tau)^n \Gamma_h (\omega+n\tau) \Gamma_h \left(2\omega-m-\frac{\tau}{2} -n\tau\right)}
			{(-\omega_1 \omega_2)^\frac{n}{2} 2^{n} n!}
			\int
			\prod_{i=1}^{n} dx_i
			\frac{ \Gamma_h\left(  \pm x_i +m+ \frac{\tau}{2}\right) \Gamma_h\left(m+\frac{\tau}{2}\right)}
			{\Gamma_h(\pm x_i)}
			\\&\times
			\prod_{1\leq i<j \leq n}\frac{\Gamma_h\left(  \pm x_i \pm x_j + \tau\right)
				\Gamma_h(\pm x_i +\tau)}
			{ \Gamma_h(\pm x_i \pm x_j)} 
		\end{split}
	\end{equation}
	The  mass parameters for the symplectic gauge group satisfy
	\begin{equation}
		2\omega - (2n+1)\frac{\tau}{2} + \sum_{i=1}^{n}\left(\pm x_i +\frac{\tau}{2}\right) + \frac{\tau}{2} = 2\omega
	\end{equation}
	We then deconfine the adjoint using the confining duality with linear monopole superpotential \eqref{USp}:\\
	\tikzmark{T2'}
	\begin{equation}
	\tikzmark{T2'}
		\begin{split}
			Z_{\mathcal{T}_{2'}}
			=&
			\frac{ \Gamma_h(\tau)^n \Gamma_h (\omega + n\tau) \Gamma_h \left(2\omega-m-\frac{\tau}{2} -n\tau\right)}
			{(-\omega_1 \omega_2)^\frac{n(n-1)}{2} 2^{n(n-1)} n! (n-1)!}
			\int
			\prod_{i=1}^{n} dx_i
			\frac{ \Gamma_h\left(  \pm x_i +m+ \frac{\tau}{2}\right) \Gamma_h\left(m+\frac{\tau}{2}\right)}
			{\Gamma_h(\pm x_i)}
			\\&\times
			\prod_{1\leq i<j \leq n}\frac{\Gamma_h(\pm x_i + n\tau)\Gamma_h(n\tau)}
			{ \Gamma_h(\pm x_i \pm x_j)} 
			\prod_{a=1}^{n-1}  dy_a
			\frac{\Gamma_h \left(\pm y_a + 2\omega -(2n+1)\frac{\tau}{2}\right)}
			{\Gamma_h (\pm 2y_a)}
			\\&\times
			\prod_{1\leq a<b \leq n-1}\frac{1}{ \Gamma_h(\pm y_a \pm y_b)} 
			\prod_{a=1}^{n-1} \prod_{i=1}^{n} \Gamma_h\left(\pm x_i \pm y_a + \frac{\tau}{2}\right)
			\		\Gamma_h\left(\pm y_a + \frac{\tau}{2}\right)
		\end{split}	
		\label{eq:BLM_T2prime}
	\end{equation}
	The last step consists in dualising the orthogonal node with \eqref{eq:aharony_SO_odd_2}:\\
	\tikzmark{T3}
	\begin{equation}
		\begin{split}
			Z_{\mathcal{T}_3}	\label{eq:T3_BLM}
			=&
			\frac{ \Gamma_h(\tau)^n \Gamma_h\left(\omega -m-\frac{\tau}{2}-(2n-1)\tau\right)
				\Gamma_h(2m+\tau) \Gamma_h(2n\tau) \Gamma_h\left( \omega-m-\frac{\tau}{2} \right)
			}
			{(-\omega_1 \omega_2)^\frac{n-1}{2} 2^{n-1} (n-1)!}
			\\&\times
			\int
			\prod_{a=1}^{n-1} dy_a  
			\frac{ \Gamma_h (\pm y_a+ m+\tau) \Gamma_h\left(\pm y_a + \omega - \frac{\tau}{2}\right) }
			{\Gamma_h(\pm 2 y_a)}
			\prod_{1\leq a<b \leq n}\frac{ \Gamma_h(\pm y_a \pm y_b+\tau )}{ \Gamma_h(\pm y_a \pm y_b)} 
		\end{split}
	\end{equation}	
	This is equivalent to the theory $\mathcal{T}_1$ with a lower rank and additional singlets. The new mass for the fundamental $q$ is $\tilde{m}=m+\tau$. The whole step is shown schematically in Figure \ref{fig:dec_quiver_4}.
By iterating these steps $n$ times one gets to a confining theory with singlets described by \eqref{lastuspblm}.

\subsubsection*{Case II: $SO(2n)$}
	Now we consider a $SO(2n)$ theory with one fundamental and one adjoint with $W=0$. The partition function is:
	\begin{equation}
		\frac{ \Gamma_h(\tau)^n }
		{(-\omega_1 \omega_2)^\frac{n-1}{2} 2^{n} n!}
		\int
		\prod_{i=1}^{n} dx_i
		\Gamma_h\left(  \pm x_i +m\right)
		\prod_{1\leq i<j \leq n}\frac{\Gamma_h\left(  \pm x_i \pm x_j + \tau\right)}
		{ \Gamma_h(\pm x_i \pm x_j)} 
	\end{equation}
	we deconfine the adjoint with \eqref{eq:aharony_USp} and get to a quiver with gauge groups $USp(2n-2)\times SO(2n)$:
	
	\begin{equation}
		\begin{split}
			&\frac{ \Gamma_h(\tau)^n \Gamma_h (n\tau)}
			{(-\omega_1 \omega_2)^{n-1} 2^{n(n-1)} n!(n-1)!}
			\int
			\prod_{i=1}^{n} dx_i
			\Gamma_h\left(  \pm x_i +m\right)
			\prod_{1\leq i<j \leq n}\frac{1}
			{ \Gamma_h(\pm x_i \pm x_j)} 
			\\&\times
			\prod_{a=1}^{n-1} dy_a 
			\frac{ \Gamma_h \left( \pm x_i \pm y_a + \frac{\tau}{2} \right)}
			{\Gamma_h (\pm 2 y_a)}
			\prod_{1\leq a < b \leq n-1} \frac{1} {\Gamma_h (\pm y_a\pm y_b)}			
		\end{split}
	\end{equation}
	
	Next we dualise the orthogonal node:
	\begin{equation}
		\begin{split}
			&\frac{ \Gamma_h(\tau)^n  \Gamma_h (n\tau) \Gamma_h (\omega-(n-1)\tau - m) \Gamma_h(2m)
				\Gamma_h(\omega-m)}
			{(-\omega_1 \omega_2)^\frac{n-1}{2} 2^{n-1} (n-1)!}
			\\&\times
			\int
			\prod_{a=1}^{n-1} dy_a 
			\frac{ \Gamma_h \left( \pm y_a + \omega - \frac{\tau}{2} \right) \Gamma_h \left(\pm y_a +m+\frac{\tau}{2}\right)}
			{\Gamma_h (\pm 2 y_a)}
			\prod_{1\leq a < b \leq n-1} \frac{\Gamma_h (\pm y_a\pm y_b+\tau)}
			{\Gamma_h (\pm y_a\pm y_b)}		
		\end{split}
	\end{equation}
	This is the $USp$ theory with adjoint considered in the previous case with additional singlets. We use the result from the previous case to confine the gauge theory and recover \eqref{risSO}.

\subsubsection*{Case III: $SO(2n+1)$}
	The case of orthogonal gauge group with odd rank is already covered in the computation for symplectic gauge group. This theory corresponds to the third step in the $USp$ computation, namely \eqref{eq:T2_BLM}, modulo the presence of some singlets. One can follow the confinement/deconfinement steps going from  \eqref{eq:T2_BLM} to \eqref{eq:T3_BLM}, then confine the $USp$ gauge theory using the result from the previous case.

%
\section{$SO(N)$ with $N+1$ flavors and linear monopole superpotential}
\label{sec:B}
%
%
In this appendix we review the duality for $SO(N)$ gauge theories with 
$N+1$ vectors $Q_i$ and $W = Y_+$ proposed by \cite{Benvenuti:2021nwt}.
We further discuss the related identity between the partition functions. This is useful for the proofs of the dualities in the body of the paper because we use such dualities to deconfine the adjoint of symplectic gauge groups.

In this case the claim is that the model is dual to a WZ model, where the fields are the baryons $q=\epsilon_N (Q^N)$ and the symmetric meson $S$ with superpotential $W = q S q + \det S$.
In order to obtain the partition function for such a duality we start from
$USp(2n)$ with linear monopole superpotential $W = Y_{USp}$
and $2n+4$ fundamentals. The linear monopole imposes the constraint
$ \mu_1+\dots+\mu_{2n+4}=2 \omega$ on the mass parameters $\mu_r$ of the fundamental fields in the partition function.
The integral identity is \cite{VanDeBult}
\begin{eqnarray}
\label{USp}
&&
\frac{1}{(-\omega_1 \omega_2)^{\frac{n}{2}}2^n n!}
\int_{C^n}\prod_{1\leq j<k \leq n} \frac{1}{\Gamma_h(\pm x_j \pm x_k)}
 \prod_{j=1}^{n}  \frac{\prod_{r=1}^{2n+4} \Gamma_h(\mu_r\pm x_j)}{\Gamma_h(\pm 2x_j)} d x_j 
\nonumber \\
&&=
 \prod_{1 \leq r < s \leq 2n+4}
\Gamma_h(\mu_r+\mu_s)
\end{eqnarray} 
If we then assign the  mass parameters as
$\mu_1 = \frac{\omega_1}{2}$ and 
$\mu_2 = \frac{\omega_1}{2}$, and we use the duplication formula
 on both sides of (\ref{USp}), 
we arrive at the identity
 \begin{eqnarray}
\label{SOdd}
&&
\frac{\prod_{r=1}^{2n+2} \Gamma_h(\mu_r)} {(-\omega_1 \omega_2)^{\frac{n}{2}}2^n n!}
\int_{C^n}\prod_{1\leq j<k \leq n} \frac{1}{\Gamma_h(\pm x_j \pm x_k)}
 \prod_{j=1}^{n}  \frac{\prod_{r=1}^{2n+2} \Gamma_h(\mu_r\pm x_j)}{\Gamma_h(\pm x_j)} d x_j 
\nonumber \\
&&=
 \prod_{1 \leq r \leq s \leq 2n+2}
\Gamma_h(\mu_r+\mu_s)
\prod_{r=1}^{2n+2} \Gamma_h(\omega-\mu_r)
\end{eqnarray} 
with the constraint $\sum_{r=1}^{2n+2} \mu_r= \omega$.
This corresponds to the case of $SO(2n+1)$ with $2n+2$ fundamentals.
The arguments of the singlets on the dual side correspond to the mesons and to the baryons of the electric theory.

The even case is obtained by considering also $\mu_3=0$.
In this case, by using the duplication formula on both sides of (\ref{USp}) 
 we end up with 
 \begin{eqnarray}
\label{SOeven}
&&
\frac{1}{(-\omega_1 \omega_2)^{\frac{n}{2}}2^{n-1} n!}
\int_{C^n}\prod_{1\leq j<k \leq n} \frac{1}{\Gamma_h(\pm x_j \pm x_k)}
 \prod_{j=1}^{n}  \prod_{r=1}^{2n+1} \Gamma_h(\mu_r\pm x_j)
 d x_j 
\nonumber \\
&&=
 \prod_{1 \leq r \leq s\leq 2n+1}
\Gamma_h(\mu_r+\mu_s)
\prod_{r=1}^{2n+1} \Gamma_h(\omega-\mu_r)
\end{eqnarray} 
with the constraint $\sum_{r=1}^{2n+1}  \mu_r= \omega$.
This corresponds to the case of $SO(2n)$ with $2n+1$ fundamentals.
The arguments of the singlets on the dual side correspond to the mesons and to the baryons of the electric theory.\\

As a consistency check we can perform a real mass flow by giving large masses of opposite sign to two vectors and retrieve the limiting case of Aharony duality. In \eqref{SOdd} we fix:
\begin{equation}
	\mu_{2n+1} = s+\nu, \qquad \mu_{2n+2} = -s+\nu
\end{equation}
and take the limit $s\to\infty$. The constraint reads $\omega - \sum_{r=1}^{2n} \mu_r = 2\nu$ and the divergent phases cancel between the RHS and the LHS. We obtain:
\begin{eqnarray}
&&
\frac{\prod_{r=1}^{2n} \Gamma_h(\mu_r)} {(-\omega_1 \omega_2)^{\frac{n}{2}}2^n n!}
\int_{C^n}\prod_{1\leq j<k \leq n} \frac{1}{\Gamma_h(\pm x_j \pm x_k)}
 \prod_{j=1}^{n}  \frac{\prod_{r=1}^{2n} \Gamma_h(\mu_r\pm x_j)}{\Gamma_h(\pm x_j)} d x_j 
\nonumber \\
&&=
\Gamma_h\left(\omega - \sum_{r=1}^{2n} \mu_r\right)
 \prod_{1 \leq \mu_r \leq \mu_s\leq 2n}
\Gamma_h(\mu_r+\mu_s)
\prod_{r=1}^{2n} \Gamma_h(\omega-\mu_r)
\end{eqnarray} 
which corresponds to the limiting case of Aharony duality for $SO(N) = SO(2n+1)$ and $ 2n$ vectors, with $W=0$ \cite{Benini:2011mf}. 

Similarly in \eqref{SOeven} we fix:
\begin{equation}
	\mu_{2n} = s+\nu, \qquad \mu_{2n+1} = -s+\nu
\end{equation}
and obtain:
\begin{eqnarray}
&&
\frac{1}{(-\omega_1 \omega_2)^{\frac{n}{2}}2^{n-1} n!}
\int_{C^n}\prod_{1\leq j<k \leq n} \frac{1}{\Gamma_h(\pm x_j \pm x_k)}
 \prod_{j=1}^{n}  \prod_{r=1}^{2n-1} \Gamma_h(\mu_r\pm x_j)
 d x_j 
\nonumber \\
&&=
\Gamma_h\left(\omega - \sum_{r=1}^{2n-1} \mu_r\right)
 \prod_{1 \leq \mu_r \leq \mu_s\leq 2n-1}
\Gamma_h(\mu_r+\mu_s)
\prod_{r=1}^{2n-1} \Gamma_h(\omega-\mu_r)
\end{eqnarray} 
which corresponds to the limiting case of Aharony duality for $SO(N) = SO(2n)$ and $ 2n-1$ vectors, with $W=0$ \cite{Benini:2011mf}.


\bibliographystyle{JHEP}
\bibliography{ref}

\end{document}